\documentclass[preprint,showpacs,preprintnumbers,amsmath,amssymb]{revtex4}

\usepackage{graphicx}    
\usepackage{dcolumn}     
\usepackage{bm}          

\begin{document}

\preprint{}
\title{Theory of charge transport in diffusive normal metal / unconventional
singlet superconductor  contacts }
\author{Y. Tanaka$^{1,2,3}$, Yu. V. Nazarov$^3$, A.A. Golubov $^4$, and S.
Kashiwaya$^5$}
\affiliation{$^1$Department of Applied Physics, Nagoya University, Nagoya, 464-8603,
Japan \\
$^2$CREST Japan Science and Technology Cooperation (JST) Nagoya 
464-8603, Japan  \\
$^3$Department of Nanoscience, Faculty of Applied Sciences, Delft
University of Technology, 2628 CJ Delft The Netherlands \\
$^4$Faculty of Science and Technology, University of Twente, 7500
AE, Enschede,
The Netherlands \\
$^5$National Institute of Advanced Industrial Science and Technology,
Tsukuba, 305-8568, Japan }
\date{\today}

\begin{abstract}
We analyze the transport properties of  contacts between unconventional
superconductor and normal diffusive
metal  in the framework of the extended circuit theory.
We obtain a general boundary condition for the Keldysh-Nambu Green's functions
at the interface that is valid for arbitrary transparencies of
the interface. This allows us to investigate
the voltage-dependent conductance (conductance spectrum)
of a diffusive normal metal (DN)/ unconventional singlet  superconductor
junction in both ballistic and diffusive cases.
For $d$-wave superconductor, we calculate conductance
spectra  numerically for different
orientations of the junctions,
 resistances, Thouless energies in DN, and
transparencies of the interface.
We demonstrate that conductance spectra exhibit
a variety of features
including a $V$-shaped gap-like
structure, zero bias conductance peak (ZBCP) and zero bias conductance
dip (ZBCD).
We show that two distinct mechanisms:
(i) coherent Andreev reflection (CAR) in DN and
(ii) formation of midgap Andreev bound state (MABS) at
the interface of $d$-wave superconductors,
are responsible for ZBCP,
their relative importance being dependent on the angle $\alpha$
between the interface normal and the crystal axis of $d$-wave
superconductors. For $\alpha=0$, the ZBCP is due to CAR
in the junctions of low transparency with small
Thouless energies, this is similar to the case of
diffusive normal metal / insulator /s-wave
superconductor junctions.
With increase of $\alpha$ from zero to $\pi/4$,
the MABS contribution to ZBCP becomes more
prominent and the effect of CAR
is gradually suppressed.
Such complex spectral features shall be observable
in conductance spectra of realistic high-$T_c$ junctions
at very low temperature.
\end{abstract}

\pacs{PACS numbers: 74.20.Rp, 74.50.+r, 74.70.Kn}
\maketitle



%

%




\section{Introduction}

The low energy transport in mesoscopic superconducting
systems is governed by Andreev reflection \cite{Andreev},
a unique process specific to electron scattering at
normal metal/superconductor interfaces. The
phase coherence between incoming electrons and Andreev reflected holes
persists at a mesoscopic length scale in the diffusive normal metal, which
enhances interference effects on the probability of Andreev reflection
\cite{Hekking}. The coherence plays an important role at
sufficiently low temperatures and voltages
when the energy broadening due to either voltage or temperature becomes
of the order of the Thouless energy $E_{Th}$ of the mesoscopic structure.
As a result, the
conductance spectra of mesoscopic junctions may be significantly
modified by these interference effects.
A remarkable experimental manifestation of the electron-hole
phase coherence is the observation of the zero bias conductance peak (ZBCP)
in diffusive normal metal (N) / superconductor (S) tunneling
junctions \cite%
{Giazotto,Klapwijk,Kastalsky,Nguyen,Wees,Nitta,Bakker,Xiong,Magnee,Kutch,Poirier}%
.

Various theoretical models of charge transport in diffusive junctions
extend the clean limit theories developed by Blonder,
Tinkham and Klapwijk \cite{BTK} (BTK) and Zaitsev. \cite{Zaitsev} In Refs.
\cite{Beenakker1,Beenakker2,Lambert,Takane,reflec,Lesovik} the scattering
matrix approach was used. On the other hand, the quasiclassical Green's
function method in nonequilibrium superconductivity \cite%
{Larkin} is much more powerful and convenient for the actual calculations of
conductance for the arbitrary bias-voltages \cite{Volkov}. Using the
Kuprianov and Lukichev (KL) boundary condition \cite{KL}  for a
diffusive SIN interface, Volkov,
Zaitsev and Klapwijk (VZK) have obtained the conductance spectra
with ZBCP, origin of which  was attributed to coherent Andreev reflection 
(CAR) which induces the proximity effect in diffusive metal
\cite{Volkov}. Several authors studied the charge transport
in mesoscopic junctions combining this boundary condition
with  Usadel \cite{Usadel} equations that describe superconducting correlations
in diffusive metal \cite%
{Nazarov1,Yip,Stoof,Reentrance,Golubov,Takayanagi,Seviour,Belzig}.

The modified boundary conditions were studied by
several authors \cite{Lambert1,Bezuglyi}. Important progress was
achieved by one of the authors  \cite{Nazarov1,Nazarov2} who developed
the so-called "circuit theory" for matrix currents that allows to
formulate boundary conditions for Usadel-like equations
in the case of arbitrary transparencies.
By using this generalized boundary condition, three of
the authors have evaluated the conductance in DN/S junctions and
demonstrated how various may be the conductance spectra.\cite{Golubov2003}.

Another mechanism of ZBCP plays the role
in ballistic unconventional superconductor junctions.
The conductance peak in this situation arises from the formation of
midgap Andreev bound states (MABS) at the interface \cite%
{Buch,TK95,96,Kashi00}. The experimental observation of the ZBCP
has been reported for various unconventional superconductors of
anisotropic pairing symmetry
\cite{Kashi00,e1,e2,e3,e4,e5,e6,e7,e8,e9,e10,e11,e12,e13,e14,e15}.
A basic theory of ballistic transport in the presence of MABS has
been formulated in Refs. \cite{TK95,Kashi00}. Stimulated by this
theory, extensive studies of MABS in unconventional superconductor
junctions have been performed during the last decade: in the case
of broken time reversal symmetry state \cite
{T1,T2,T3,T4,T5,T6,T7,T8}, in triplet superconductor junctions \cite%
{tr1,tr2,tr3,tr4,tr5,tr6}, in quasi-one dimensional organic
superconductors \cite {or1,or2,or3}, MABS and Doppler effect
\cite{dop1,dop2,dop3,dop4,dop5}, MABS in ferromagnet junctions
\cite{f1,f2,f3,f4,f5,f6,f7,f8,f9}, influence of MABS on Josephson
effect \cite{TKJ,J1,J2,J3,J4,J5,J6,J7,J8,J9} and other related
problems \cite{o1,o2,o3,o4,o5,o6,o7,o8,o9}. However, an
interesting question remained: how two mechanisms of ZBCP (due to
CAR and due to MARBS) work together,  this being relevant for
diffusive normal metal / unconventional superconductor (DN/US)
junctions.

To solve this problem, three of the present authors have recently
extended the circuit theory to the systems that contain
unconventional singlet superconductor junctions
\cite{Nazarov2003}. Application of this theory for DN/$d$-wave
superconductor (DN/d) junctions has shown that the formation of
MABS strongly competes with the proximity effect that is an
essential ingredient for CAR in DN. The MABS induces the
unconventional channels where quasiparticles are resonantly
transmitted through the interface. The overall contribution of
these channels to the proximity effect is however suppressed by
the isotropization, $i.e.$, the angular averaging over momentum
directions of injected quasiparticles. However, the reference
\cite{Nazarov2003} does not contain the necessary technical
details of the matrix current derivation and presents the results
only for low voltage limit. To compare with experiment, one has to
evaluate the conductance spectrum in wide range of bias voltage.

In this paper, we present a detailed derivation of the matrix current
in  (DN/US) junctions.
Although the  relation  obtained  is valid for both
singlet and triplet superconductor junctions, we focus on the
case of singlet superconductor. We present detailed
numerical calculations of the conductance spectra of DN/US
junctions for $d$-wave
superconductors. We investigate the dependence of the
spectra on various parameters: the height of the barrier at the
interface, resistance $R_{d}$ in DN, the Thouless energy $E_{Th}$ in DN and
the angle between the normal to the interface and the crystal axis of $d$%
-wave superconductor ($\alpha $). We  normalize  the voltage-dependent
conductance $\sigma_{S}(eV)$
by its value in the normal state, $\sigma_N$,
so that $\sigma_{T}(eV)=\sigma_{S}(eV)/\sigma_N$.

Our main results are as follows:

\noindent 1. The ZBCP is frequently seen in the shape of $%
\sigma_{T}(eV)$. For $\alpha \neq 0$, the ZBCP is robust
not depending on the diffusive resistance $R_{d}$.
For $\alpha=0$, ZBCP is due to the CAR.

\noindent 2. The appearance of ZBCP
is different for MABS and CAR mechanisms.
The first mechanism may lead to arbitrarily large $\sigma _{T}(0)$.
The second mechanism can not provide $\sigma_{T} (0)$ exceeding  unity.
While for the first mechanism the width of the ZBCP is
determined by the transparency of the junction, it is determined
by  Thouless energy
for the second one.
These two mechanisms compete
since the proximity effect and the MABS in singlet junctions
are generally incompatible\cite{Nazarov2003}.

\noindent 3. In the extreme case $\alpha=\pi/4$ the proximity
effect and the CAR are absent. The $\sigma_{T}(eV)$ is then given by a simple
Ohm's law:$%
\sigma_{T}(eV)=(R_{b} + R_{d})/(R_{R_{d}=0} + R_{d})$  $R_b$ being
the resistance of the interface. \par
\noindent 4. For $\alpha =0$, when MABS are absent for $%
R_{d}=0$, the ZBCP of $\sigma _{T}(eV)$ is attributed to 
the CAR alone. When the transparency of the junction
is sufficiently low, $\sigma _{T}(eV)$ for $\mid eV\mid <\Delta _{0}$ is
enhanced with the increase of $R_{d}$ due to the enhancement of the
proximity effect ($\Delta _{0}$ is the maximum amplitude of the pair
potential). The ZBCP becomes prominent for $E_{Th}\ll\Delta _{0}$ and $%
R_{d}/R_{b}<1$ ($R_{d}\neq 0$). In this case,   the ZBCP turns into a zero bias conductance dip (ZBCD)
with a further increase of $%
R_{d}/R_{b}$ \cite{minigap}.\par
\noindent 5. We have shown that the conductance spectrum
can vary substantially depending on the parameters.
The results obtained are important
to analyze the actual experimental data on
conductance spectra of high
$T_{C}$ cuprate junctions
since in this case the diffusive scattering in the normal metal is
of special relevance.

The structure of the paper is as follows. We formulate the model
in use in section 2. We also present there the detailed derivation
of the matrix current and end up with the expression for the
normalized conductance. We focus on $d$-wave superconductor
junctions in section 3 and
evaluate $\sigma _{T}(eV)$ and the measure of the proximity effect $%
\theta_{0}$ for various cases. We summarize the results in section 4.

\section{Formulation}

In this section we introduce the model and the formalism. We consider a
junction consisting of normal and superconducting reservoirs connected by a
quasi-one-dimensional diffusive conductor (DN) with a length $L$ much larger
than the mean free path. The interface between the DN conductor and the US
(unconventional superconductor) electrode has a resistance $R_{b}$ while the
DN/N interface has zero resistance. The positions of the DN/N interface and
the DN/S interface are denoted as $x=-L$ and $x=0$, respectively. According
to the circuit theory \cite{Nazarov2}, the constriction area ($-L_{1}<x<L_{1}
$) between DN and US is considered as composed of the diffusive
isotropization zone $(-L_{1}<x<-L_{2})$, the left side ballistic zone $%
(-L_{2}<x<0)$, the right side ballistic zone $(0<x<L_{1})$ and the
scattering zone $(x=0)$. The scattering zone is modeled as an insulating
delta-function barrier with the transparency $T_{n}=4\cos^{2}\phi /(4\cos
^{2}\phi +Z^{2})$, where $Z$ is a dimensionless constant, $\phi $ is the
injection angle measured from the interface normal to the junction and $n$
is the channel index. We assume that the sizes of the ballistic and
scattering zones along $x$ axis is much shorter than the superconducting
coherence length.

%
Here, we express insulating barrier as a delta function model $H\delta (x)$,
where $Z$ is given by $Z=2mH/(\hbar ^{2}k_{F})$ with Fermi momentum $k_{F}$
and effective mass $m$.
In order to clarify charge transport in DN/US junctions, we must
obtain Keldysh-Nambu Green's function, which has indices of
transport channels and the direction of motion along $x$ axis
taking into account the proper boundary conditions. For this
purpose it is necessary to extend a general theory of boundary
condition which covers the crossover from ballistic to diffusive
cases \cite{Nazarov2} formulated for conventional junctions in the
framework of the circuit theory \cite{Nazarov1,Nazarov2}. However,
the circuit theory cannot be directly applied to unconventional
superconductors since it
requires the isotropization. %
%
To avoid this difficulty we restrict the discussion to a conventional model
of \textit{smooth interface} by assuming momentum conservation in the plane
of the interface.

In this section, we will show how to derive the matrix current in DN/US
junctions. Then we will derive\ the retarded and Keldysh components of the
matrix current. Finally, we will show how to calculate  conductance of the junctions.

\subsection{Calculation of the matrix current}

To derive the relation between the matrix current and Green's
functions, we make use of the method proposed in \cite{Nazarov2}.
The method puts the older ideas \cite{Zaitsev} to the framework of
Landauer-B\"{u}ttiker scattering formalism. One expresses the
matrix current in a constriction in terms of one-dimensional
Green's functions $\check{g}_{n,\sigma ;n^{\prime },\sigma
^{\prime }}(\varepsilon ;x,x^{\prime })$, where $n$,$n^{\prime }$
and $\sigma ,\sigma ^{\prime }=\pm 1$ denote the indices of
transport channels and the direction of motion along $x$ axis,
respectively. The "check" represents the Keldysh-Nambu structure.
These Green's functions have to be expressed in terms of the
transfer matrix that incorporates all information about the
scattering, and asymptotic Green's functions presenting boundary
conditions deep in each side of the constriction. Here, we
restrict the discussion to a conventional model of \textit{smooth
interface}, assuming momentum conservation in the plane of the
interface. Within the model, the channel number eventually numbers
possible values of this in-plane momentum and the transfer matrix
becomes block-diagonal in the channel index. Following the
treatment developed in Ref. \cite{Nazarov2}, we thus solve Green's
functions $\check{g}_{n,\sigma ;n^{\prime },\sigma
^{\prime }}(\varepsilon ,x,x^{\prime })$ separately for each channel. $%
\check{g}_{n,\sigma ;n^{\prime },\sigma ^{\prime }}(\varepsilon ,x,x^{\prime
})$ can be expressed as
\begin{equation}
\check{g}_{n,\sigma ;n^{\prime },\sigma ^{\prime }}(\varepsilon ,x,x^{\prime
})=\sum_{\sigma ,\sigma ^{\prime }=\pm 1}\exp (i\sigma p_{n}x-i\sigma
^{\prime }p_{n}x^{\prime })\check{G}_{n}^{\sigma ,\sigma ^{\prime
}}(x,x^{\prime })
\end{equation}

The function $\check{G}^{\sigma,\sigma^{\prime}}_{n}(x,x^{\prime})$ are
varying smoothly at the scale of $1/p$ and obey the following semi classical
equation.

\begin{equation}
(i\sigma v_{n} \frac{\partial }{\partial x} + \check{H}(x)) \check{G}%
^{\sigma,\sigma^{\prime}}_{n}(x,x^{\prime})=0
\end{equation}
with effective Hamiltonian $\check{H}(x)$ where $\check{H}(x)$ and $\check{H}%
_{0}(x)$ are given by
\begin{equation}
\check{H}(x) =\check{H}_{0}(x) - \check{\Sigma}_{imp}(x),
\end{equation}
\begin{equation}
\check{H}_{0}(x)=\varepsilon \check{\tau}_{z} +\check{\Delta}(x), \ \ \check{%
\tau}_{z} = \left(
\begin{array}{cc}
\hat{\tau}_{z} & 0 \\
0 & \hat{\tau}_{z}%
\end{array}
\right).
\end{equation}
In the above, $\check{\Sigma}_{imp}(x)$ is a self-energy due to the impurity
scattering and $\check{\Sigma}_{imp}(x) \neq 0$ is satisfied only for $x<0$.
The self-energy originating from the superconducting pair potential $\check{%
\Delta}(x)=0$ for $x<0$ while $\check{\Delta}(x)$ for $x>0$ is given by
\begin{equation}
\\
\check{\Delta}(x) = \left(
\begin{array}{cc}
\hat{\Delta}_{\sigma} & 0 \\
0 & \hat{\Delta}_{\sigma}%
\end{array}
\right), \\
\hat{\Delta}_{\sigma} = \left(
\begin{array}{cc}
0 & \Delta_{\sigma} \\
-\Delta^{*}_{\sigma} & 0%
\end{array}
\right).
\end{equation}
$\check{G}^{\sigma\sigma^{\prime}}_{n}(x,x^{\prime})$ has discontinuity at $%
x=x^{\prime}$

\begin{equation}
\check{G}_{n}^{\sigma \sigma ^{\prime }}(x+0,x)-\check{G}_{n}^{\sigma \sigma
^{\prime }}(x-0,x)=-i\check{1}\sigma \delta _{\sigma \sigma ^{\prime }}
/\mid v_{n} \mid.
\end{equation}%
To get rid of the discontinuity, we define $\bar{g}_{n}^{\sigma \sigma ^{\prime
}}(x,x^{\prime })$ as follows
\begin{equation}
2i\check{G}_{n}^{\sigma \sigma ^{\prime }}(x,x^{\prime })=[\bar{g}%
_{n}^{\sigma \sigma ^{\prime }}(x,x^{\prime })+\sigma \mathrm{sign}%
(x-x^{\prime })]/\mid v_{n}\mid .
\end{equation}%
We denote $\bar{g}_{n}^{\sigma \sigma ^{\prime }}(0_{+},0_{+})=\bar{g}_{2}$ and $%
\bar{g}_{n}^{\sigma \sigma ^{\prime }}(0_{-},0_{-})=\bar{g}_{1}$, where $\bar{g}%
_{2}$ and $\bar{g}_{1}$ satisfy
\[
\bar{g}_{2}=\bar{M}^{\dagger }\bar{g}_{1}\bar{M},
\]%
using a transfer matrix ${\bar{M}}$.

\bigskip The derivation of the matrix current

\begin{equation}
\check{I}=\frac{2e^{2}}{h}\mathrm{Tr}_{n,\sigma }[\bar{\Sigma}^{z}\bar{g}%
_{1}]=\frac{2e^{2}}{h}\mathrm{Tr}_{n,\sigma }[\bar{\Sigma}^{z}\bar{g}_{2}].
\end{equation}%
is given in Appendix, where it is shown that $\check{I}_{n0}$ can
be represented as follows \cite{Nazarov2003},
\[
\check{I}_{n0}=2[\check{G}_{1},\check{B}_{n}]
\]%
with
\begin{equation}
\check{B}_{n}=(-T_{1n}[\check{G}_{1},\check{H}_{-}^{-1}]+\check{H}_{-}^{-1}%
\check{H}_{+}-T_{1n}^{2}\check{G}_{1}\check{H}_{-}^{-1}\check{H}_{+}%
\check{G}_{1})^{-1}(T_{1n}(1-\check{H}_{-}^{-1})+T_{1n}^{2}\check{G}_{1}%
\check{H}_{-}^{-1}\check{H}_{+})
\end{equation}%
This is the very equation, which was firstly derived as eq. (2) of Ref. \cite%
{Nazarov2003}. It should be remarked that this formula of the matrix current
is very general since it is available both for singlet and triplet
superconductors. For low transparency limit, $i.e.$, $T_{n}<<1$, $T_{1n}<<1$%
, $\check{I}_{n0}$ can be approximated to be
\begin{equation}
\check{I}_{n0}=\frac{T_{n}}{2}[\check{H}_{+}^{-1}(1-\check{H}_{-}),\check{G}%
_{1}]  \label{KL}
\end{equation}%
Eq. (\ref{KL}) can be regarded as an extended version of Kuprianov and
Lukichev's boundary condition \cite{KL} 
for unconventional superconductor junctions.
On the other hand, when $\check{G}_{2+}=\check{G}_{2-}$ is satisfied as in
the conventional superconductor, since
\[
\mathrm{lim}_{\check{H}_{-}\rightarrow 0}-\check{D}^{-1}[T_{1n}(2\check{G}%
_{1}-[\check{H}_{-}^{-1},\check{G}_{1}]_{+})+\check{H}_{-}^{-1}\check{H}%
_{+}+T_{1n}^{2}\check{G}_{1}\check{H}_{-}^{-1}\check{H}_{+}\check{G}_{1}]
\]%
\[
=(\check{H}_{+}+T_{1n}\check{G}_{1})^{-1}(-\check{H}_{+}+T_{1n}\check{G}_{1})
\]%
and
\[
\mathrm{lim}_{\check{H}_{-}\rightarrow 0}\check{G}_{1}\check{D}^{-1}\check{G}%
_{1}[T_{1n}(2\check{G}_{1}-[\check{H}_{-}^{-1},\check{G}_{1}]_{+})+T_{1n}^{2}%
\check{H}_{-}^{-1}\check{H}_{+}+\check{G}_{1}\check{H}_{-}^{-1}\check{H}_{+}%
\check{G}_{1}]
\]%
\[
=(\check{G}_{1}+T_{1n}\check{H}_{+})^{-1}(-T_{1n}\check{H}_{+}+\check{G}_{1})
\]%
are satisfied, then the resulting $\check{I}_{n0}$ and $\check{I}$ are 
reduced to be
\begin{equation}
\check{I}_{n0}=2T_{1n}(1+T_{1n}^{2}+T_{1n}[\check{H}_{+},\check{G}%
_{1}]_{+})^{-1}[\check{H}_{+},\check{G}_{1}]
\end{equation}%
\[
\check{I}=\sum_{n}\frac{2e^{2}}{h}(1+T_{1n}^{2}+T_{1n}[\check{H}_{+},\check{G%
}_{1}]_{+})^{-1}[\check{H}_{+},\check{G}_{1}]
\]%
which is identical to eq. (36) of Ref. \cite{Nazarov2}.
In the above, the definition of $\check{D}$ is given in the
Appendix.

\subsection{Calculation of the retarded part of the matrix current}

In order to calculate the retarded part of the matrix current, we denote
Keldysh-Nambu Green's function $\check{G}_{1}$, $\check{G}_{2\pm}$,
\begin{equation}
\check{G}_{1} =\left(
\begin{array}{cc}
\hat{R}_{1} & \hat{K}_{1} \\
0 & \hat{A}_{1}%
\end{array}%
\right), \ \ \check{G}_{2\pm}=\left(
\begin{array}{cc}
\hat{R}_{2\pm} & \hat{K}_{2\pm} \\
0 & \hat{A}_{2\pm}%
\end{array}%
\right),
\end{equation}%
where the Keldysh component $\hat{K}_{1,2\pm}$ is given by $\hat{K}_{1(2\pm)}=\hat{%
R}_{1(2\pm)}\hat{f}_{1(2)}(0)-\hat{f}_{1(2)}(0)\hat{A}_{1(2\pm)}$ with the
retarded component $\hat{R}_{1,2\pm}$ and
the advanced component $\hat{A}_{1,2\pm}$
using distribution function $\hat{f}_{1(2)}(0)$. In the above, $\hat{R}_{2\pm}$
is expressed by
\[
\hat{R}_{2\pm}=(g_{\pm}\hat{\tau}_{3}+f_{\pm}\hat{\tau}_{2})
\]%
with $g_{\pm}=\varepsilon /\sqrt{\varepsilon ^{2}-\Delta _{\pm}^{2}}$, $%
f_{\pm}=\Delta _{\pm}/\sqrt{\Delta _{\pm}^{2}-\varepsilon ^{2}}$, and $\hat{A%
}_{2\pm}=-\hat{\tau}_{3}\hat{R}_{2\pm}^{\dagger }\hat{\tau}_{3}$ where $%
\varepsilon $ denotes the quasiparticle energy measured from the Fermi
energy. $\hat{f}_{2}(0)=f_{0S}(0)=\mathrm{{tanh}[\varepsilon /(2k_{B}T)]}$
in thermal equilibrium with temperature $T$. Here, we put the electrical
potential zero in the US-electrode. We also denote $\check{H}_{+}$, $\check{H%
}_{-}$, $\check{B}_{n}$, $\check{I}$ as follows,
\begin{eqnarray}
\check{H}_{+} =\left(
\begin{array}{cc}
\hat{R}_{p} & \hat{K}_{p} \\
0 & \hat{A}_{p}%
\end{array}%
\right) , \ \ \check{H}_{-} =\left(
\begin{array}{cc}
\hat{R}_{m} & \hat{K}_{m} \\
0 & \hat{A}_{m}%
\end{array}%
\right),
\end{eqnarray}

\[
\check{B}_{n}=\left(
\begin{array}{cc}
\hat{B}_{R} & \hat{B}_{K} \\
0 & \hat{B}_{A}%
\end{array}%
\right) ,\ \ \check{I}=\left(
\begin{array}{cc}
\hat{I}_{R} & \hat{I}_{K} \\
0 & \hat{I}_{A}%
\end{array}%
\right) ,
\]%
Hereafter, in the present paper, we focus on the singlet superconductor junction case without broken time reversal symmetry states (BTRSS). For
triplet case or singlet one with BTRSS, 
the situation becomes much more complex \cite{triplet2003} and
we will discuss in forthcoming paper in detail. In singlet superconductors,
we can choose $\hat{R}_{1}=\cos \theta _{0}\hat{\tau}_{3}+\sin \theta _{0}%
\hat{\tau}_{2}$ to satisfy the boundary condition at the interface. After
some algebra, we can obtain $\hat{B}_{R}$ as follows
\[
\hat{B}_{R}=-T_{1n}[1+T_{1n}^{2}+T_{1n}(\hat{R}_{1}\hat{R}_{p}^{-1}+\hat{R}%
_{p}^{-1}R_{1})]^{-1}[T_{1n}R_{1} + \hat{R}_{p}^{-1}]
\]%
where we have used the relation
\[
\hat{R}_{1}\hat{R}_{m}^{-1}\hat{R}_{p}+\hat{R}_{m}^{-1}\hat{R}_{p}\hat{R}%
_{1}=0
\]%
The resulting $\hat{I}_{R}$ can be written as
\begin{equation}
\hat{I}_{R}=\frac{2e^{2}}{h}\sum_{n}\frac{-2T_{n}[\cos \theta
_{0}(f_{+}+f_{-})-\sin \theta _{0}(g_{+}+g_{-})]}{%
(2-T_{n})(1+g_{+}g_{-}+f_{+}f_{-})+T_{n}[\cos \theta _{0}(g_{+}+g_{-})+\sin
\theta _{0}(f_{+}+f_{-})]}\hat{\tau}_{3}\hat{\tau}_{2}
\end{equation}%
This is one of the central results of this paper.

\subsection{Calculation of the Keldysh part of the matrix current}

Next, we focus on the Keldysh component. We define $I_{b}$
\begin{equation}
I_{b}=\frac{e^{2}}{h}\sum_{n}\mathrm{Tr}[\hat{I}_{K}\hat{\tau}_{3}].
\end{equation}%
After straightforward calculations, $I_{b}$ is given by
\[
I_{b}=\frac{e^{2}}{h}\sum_{n}\mathrm{Trace}\{\hat{\tau}_{3}(\hat{R}_{1}\hat{B%
}_{K}+\hat{R}_{1}^{\dagger }\hat{B}_{K})
\]%
\[
-[\hat{\tau}_{3}(\hat{R}_{1}^{\dagger }\hat{B}_{R}^{\dagger }+\hat{B}_{R}%
\hat{R}_{1}+\hat{B}_{R}^{\dagger }\hat{R}_{1}+\hat{R}_{1}^{\dagger }\hat{B}%
_{R})]f_{0N}(0)-[(\hat{R}_{1}+\hat{R}_{1}^{\dagger })(\hat{B}_{R}+\hat{B}%
_{R}^{\dagger })]f_{3N}(0)\},
\]%
with $\hat{K}_{1}=\hat{R}_{1}\hat{f}_{1}(0)-\hat{f}_{1}(0)\hat{A}_{1}$ $%
\hat{f}_{1}(0)=f_{0N}(0)+f_{3N}(0)\hat{\tau}_{3}$. Since $\hat{B}_{R}$ and $%
\hat{R}_{1}$ is proportional to the linear combination of $\hat{\tau}_{2}$
and $\hat{\tau}_{3}$, the second term which is proportional to $f_{0N}(0)$
disappears. It is necessary to obtain $\hat{B}_{K}$ which is given by
\begin{equation}
\hat{B}_{K}=\hat{D}_{R}^{-1}\hat{N}_{K}-\hat{D}_{R}^{-1}\hat{D}_{K}\hat{D}_{A}^{-1}\hat{%
N}_{A},
\end{equation}
with
\[
\check{D}= -T_{1n}[\check{G}_{1},\check{H}_{-}^{-1}] +
\check{H}_{-}^{-1}\check{H}_{+}
-T_{1n}^{2}\check{G}_{1}\check{H}_{-}^{-1}\check{H}_{+}\check{G}_{1}, \ \
\check{D}=\left(
\begin{array}{cc}
\hat{D}_{R} & \hat{D}_{K} \\
0 & \hat{D}_{A}%
\end{array}%
\right),
\]
where $\hat{N}_{K}$ and $\hat{N}_{A}$ is the Keldysh and advanced part of
$\check{N}$ given by
\[
\check{N}=-T_{1n}\check{H}_{-}^{-1}+T_{1n}^{2}\check{G}_{1}\check{H}_{-}^{-1}%
\check{H}_{+} \ \
\check{N}=\left(
\begin{array}{cc}
\hat{N}_{R} & \hat{N}_{K} \\
0 & \hat{N}_{A}%
\end{array}%
\right).
\]%
We can express $\hat{N}_{K}$ and $\hat{D}_{K}$ as linear combination of
distribution functions $f_{0S}(0)$, $f_{0N}(0)$, and $f_{3N}(0)$ as follows,
\[
\hat{N}_{K}=\hat{C}_{1}f_{0S}(0)+\hat{C}_{2}f_{0N}(0)+\hat{C}_{3}f_{3N}(0),
\]%
\[
\hat{D}_{K}=\hat{C}_{4}f_{0S}(0)+\hat{C}_{5}f_{0N}(0)+\hat{C}_{6}f_{3N}(0),
\]%
by $2\times 2$ matrix $C_{i}$ $(i=1..6)$. Taking account of the fact that $%
\hat{D}_{R}^{-1}\hat{C}_{1}$, $\hat{D}_{R}^{-1}\hat{C}_{2}$, $\hat{D}%
_{R}^{-1}\hat{C}_{4}\hat{D}_{A}^{-1}\hat{N}_{A}$ and $\hat{D}_{R}^{-1}\hat{C}%
_{5}\hat{D}_{A}^{-1}\hat{N}_{A}$ can be expressed by the linear combination
of $\hat{\tau}_{2}$ and $\hat{\tau}_{3}$, while $\hat{D}_{R}^{-1}\hat{C}_{3}$
and $\hat{D}_{R}^{-1}\hat{C}_{6}\hat{D}_{A}^{-1}\hat{N}_{A}$ are
proportional to the linear combination of $\hat{1}$ and $\hat{\tau}_{1}$, we
can express $I_{b}$ as follows,
\[
I_{b}=\frac{e^{2}}{h}\sum_{n}\mathrm{Trace}\{(\hat{R}_{1}+\hat{R}%
_{1}^{\dagger })[\hat{B}_{KE}\hat{\tau}_{3}-(\hat{B}_{R}+\hat{B}%
_{R}^{\dagger })]f_{3N}(0)\},
\]%
\[
\hat{B}_{KE}=\hat{D}_{R}^{-1}[\hat{C}_{3}-\hat{C}_{6}\hat{D}_{A}^{-1}\hat{N}%
_{A}],
\]%
with
\[
\hat{C}_{3}=T_{1n}^{2}(\hat{R}_{1}\hat{\tau}_{3}-\hat{\tau}_{3}\hat{A}_{1})%
\hat{A}_{m}^{-1}\hat{A}_{p},
\]%
\[
\hat{C}_{6}=T_{1n}[-(\hat{R}_{1}\hat{\tau}_{3}-\hat{\tau}_{3}\hat{A}_{1})%
\hat{A}_{m}^{-1}+\hat{R}_{m}^{-1}(\hat{R}_{1}\hat{\tau}_{3}-\hat{\tau}_{3}%
\hat{A}_{1})
\]%
\begin{equation}
-T_{1n}(\hat{R}_{1}\hat{\tau}_{3}-\hat{\tau}_{3}\hat{A}_{1})\hat{A}_{m}^{-1}%
\hat{A}_{p}\hat{A}_{1}-T_{1n}\hat{R}_{1}\hat{R}_{m}^{-1}\hat{R}_{p}(\hat{R}%
_{1}\hat{\tau}_{3}-\hat{\tau}_{3}\hat{A}_{1})].
\end{equation}%
Since following equations are satisfied,
\[
\hat{D}_{A}^{-1}\hat{N}_{A}=-\tau _{3}\hat{B}_{R}^{\dagger }\tau _{3},\ \
\hat{A}_{m(p)}=-\hat{\tau}_{3}\hat{R}_{m(p)}^{\dagger }\hat{\tau}_{3},
\]%
\[
\hat{D}_{R}^{-1}(-T_{1n}\hat{R}_{m}^{-1}+T_{1n}^{2}\hat{R}_{1}\hat{R}%
_{m}^{-1}\hat{R}_{p})=\hat{B}_{R},
\]%
$I_{b}$ is given by
\[
I_{b}=\frac{e^{2}}{h}\sum_{n}[-(\hat{R}_{1}+\hat{R}_{1}^{\dagger })\hat{B}%
_{R}(\hat{R}_{1}+\hat{R}_{1}^{\dagger })\hat{B}_{R}^{\dagger }-(\hat{R}_{1}+%
\hat{R}_{1}^{\dagger })(\hat{B}_{R}+\hat{B}_{R}^{\dagger })
\]%
\[
+T_{1n}^{2}(\hat{R}_{1}+\hat{R}_{1}^{\dagger })\hat{D}_{R}^{-1}(\hat{R}_{1}+%
\hat{R}_{1}^{\dagger })(\hat{R}_{m}^{\dagger })^{-1}\hat{R}_{p}^{\dagger }
\]%
\[
+T_{1n}(\hat{R}_{1}+\hat{R}_{1}^{\dagger })\hat{D}_{R}^{-1}(\hat{R}_{1}+\hat{%
R}_{1}^{\dagger })(\hat{R}_{m}^{\dagger })^{-1}\hat{B}_{R}^{\dagger }
\]%
\begin{equation}
+T_{1n}^{2}(\hat{R}_{1}+\hat{R}_{1}^{\dagger })\hat{D}_{R}^{-1}(\hat{R}_{1}+%
\hat{R}_{1}^{\dagger })(\hat{R}_{m}^{\dagger })^{-1}\hat{R}_{p}^{\dagger }%
\hat{R}_{1}^{\dagger }\hat{B}_{R}^{\dagger }]f_{3N}(0).
\end{equation}%
After a simple manipulation, we can show
\begin{equation}
\hat{B}_{R}(\hat{R}_{m}^{-1}+T_{1n}\hat{R}_{1}\hat{R}_{p}\hat{R}%
_{m}^{-1})-T_{1n}\hat{R}_{m}^{-1}\hat{R}_{p}=-T_{1n}\hat{D}%
_{R}^{-1}=d_{R}^{-1}T_{1n}\hat{R}_{m}\hat{R}_{p}^{-1},
\end{equation}%
with
\[
d_{R}=
\frac{(1+T_{1n}^{2})(1+g_{+}g_{-}+f_{+}f_{-})
+ 2T_{1n}[\cos\theta_{0}(g_{+}+g_{-}) + \sin\theta_{0}(f_{+} + f_{-}) ] }
{1+g_{+}g_{-}+f_{+}f_{-} }.
\]
Then the resulting $I_{b}$ is given by
\[
I_{b}=\frac{e^{2}}{h}\sum_{n}[-(\hat{R}_{1}+\hat{R}_{1}^{\dagger })\hat{B}%
_{R}(\hat{R}_{1}+\hat{R}_{1}^{\dagger })\hat{B}_{R}^{\dagger }-(\hat{R}_{1}+%
\hat{R}_{1}^{\dagger })(\hat{B}_{R}+\hat{B}_{R}^{\dagger })
\]%
\begin{equation}
-T_{1n}^{2}(\hat{R}_{1}+\hat{R}_{1}^{\dagger })\hat{D}_{R}^{-1}(\hat{R}_{1}+%
\hat{R}_{1}^{\dagger })(\hat{D}_{R}^{\dagger })^{-1}]f_{3N}(0).
\end{equation}%
Since $\hat{B}_{R}$ is given as $\hat{B}_{R}=b_{2}\hat{\tau}_{2}+b_{3}\hat{%
\tau}_{3}$ with
\[
b_{2}=\frac{-T_{1n}[T_{1n}\sin \theta
_{0}(1+g_{+}g_{-}+f_{+}f_{-})+f_{+}+f_{-}]}{%
(1+T_{1n}^{2})(1+g_{+}g_{-}+f_{+}f_{-})+2T_{1n}[\cos \theta
_{0}(g_{+}+g_{-})+\sin \theta _{0}(f_{+}+f_{-})]},
\]%
\begin{equation}
b_{3}=\frac{-T_{1n}[T_{1n}\cos \theta
_{0}(1+g_{+}g_{-}+f_{+}f_{-})+g_{+}+g_{-}]}{%
(1+T_{1n}^{2})(1+g_{+}g_{-}+f_{+}f_{-})+2T_{1n}[\cos \theta
_{0}(g_{+}+g_{-})+\sin \theta _{0}(f_{+}+f_{-})]},
\end{equation}%
final expression of $I_{b}$ is given by following equation.
\begin{equation}
I_{b}=\frac{2e^{2}}{h}\sum_{n}\frac{T_{n}}{2}\frac{C_{0}f_{3N}(0)}{\mid
(2-T_{n})(1+g_{+}g_{-}+f_{+}f_{-})+T_{n}[\cos \theta _{0}(g_{+}+g_{-})+\sin
\theta _{0}(f_{+}+f_{-})]\mid ^{2}}
\end{equation}%
\[
C_{0}=T_{n}(1+\mid \cos \theta _{0}\mid ^{2}+\mid \sin \theta _{0}\mid ^{2})
\]%
\[
\times \lbrack \mid g_{+}+g_{-}\mid ^{2}+\mid f_{+}+f_{-}\mid ^{2}+\mid
1+f_{+}f_{-}+g_{+}g_{-}\mid ^{2}+\mid f_{+}g_{-}-g_{+}f_{-}\mid ^{2}]
\]%
\[
+2(2-T_{n})\mathrm{Real}\{(1+g_{+}^{\ast }g_{-}^{\ast }+f_{+}^{\ast
}f_{-}^{\ast })[(\cos \theta _{0}+\cos \theta _{0}^{\ast
})(g_{+}+g_{-})+(\sin \theta _{0}+\sin \theta _{0}^{\ast })(f_{+}+f_{-})]\}
\]%
\[
+4T_{n}\mathrm{Imag}(\cos \theta _{0}\sin \theta _{0}^{\ast })\mathrm{Imag}%
[(f_{+}+f_{-})(g_{+}^{\ast }+g_{-}^{\ast })].
\]%
This is a very general expression which is available for any singlet
superconductor without BTRSS. 
This expression is also one of the central results of this
paper. For isotropic limit where $f_{+}=f_{-}$ and $g_{+}=g_{-}$ is
satisfied, we obtain
\[
I_{b}=\frac{2e^{2}}{h}\sum_{n}\frac{T_{n}^{2}\Lambda
_{1}+2T_{n}(2-T_{n})\Lambda _{2}}{2\mid (2-T_{n})+T_{n}[g_{+}\cos \theta
_{0}+f_{+}\sin \theta _{0}]\mid ^{2}},
\]%
\[
\Lambda _{1}=(1+\mid \cos \theta _{0}\mid ^{2}+\mid \sin \theta _{0}\mid
^{2})(\mid g_{+}\mid ^{2}+\mid f_{+}\mid ^{2}+1)
\]%
\begin{equation}
+4\mathrm{Imag}[f_{+}g_{+}^{\ast }]\mathrm{Imag}[\cos \theta _{0}\sin \theta
_{0}^{\ast }],
\end{equation}%
which is identical to eq. (11) of Ref. \cite{Golubov2003}. On the other hand
for the ballistic limit, where $\theta _{0}=0$ is satisfied, we can
reproduce the generalized BTK formula \cite{TK95},
\begin{equation}
I_{b}=\frac{2e^{2}}{h}\sum_{n}\frac{T_{n}(1+T_{n}\mid \Gamma _{+}\mid
^{2}+(T_{n}-1)\mid \Gamma _{+}\Gamma _{-}\mid ^{2})}{\mid 1+(T_{n}-1)\Gamma
_{+}\Gamma _{-}\mid ^{2}}
\end{equation}%
with
\[
\Gamma _{+}=\frac{\Delta _{+} }{\varepsilon +\sqrt{\varepsilon
^{2}-\Delta _{+}^{2}}},\ \Gamma _{-}=\frac{\Delta _{-}}{\varepsilon +\sqrt{%
\varepsilon ^{2}-\Delta _{-}^{2}}}
\]%
where following equations are satisfied
\[
\frac{g_{+}+g_{-}}{1+g_{+}g_{-}+f_{+}f_{-}}=\frac{1+\Gamma _{+}\Gamma _{-}}{%
1-\Gamma _{+}\Gamma _{-}},\ \ \frac{f_{+}+f_{-}}{1+g_{+}g_{-}+f_{+}f_{-}}=%
\frac{i(\Gamma _{+}+\Gamma _{-})}{1-\Gamma _{+}\Gamma _{-}}
\]

\subsection{Calculation of the conductance}

In the following, we apply the quasiclassical Keldysh formalism for
calculation of the conductance. The spatial dependence of 4 $%
\times $ 4 Green's function in DN $\check{G}_{N}(x)$ which is expressed in
the matrix form as
\[
\check{G}_{N}(x)=\left(
\begin{array}{cc}
\hat{R}_{N}(x) & \hat{K}_{N}(x) \\
0 & \hat{A}_{N}(x)%
\end{array}%
\right) ,
\]%
should be determined. The Keldysh component $\hat{K}_{N}(x)$ is given by $%
\hat{K}_{N}(x)=\hat{R}_{N}(x)\hat{f}_{1}(x)-\hat{f}_{1}(x)\hat{A}_{N}(x)$
with retarded component $\hat{R}_{N}(x)$, advanced component $\hat{A}_{N}(x)$
using distribution function $\hat{f}_{1}(x)$.
We put the electrical potential zero in the S-electrode. In this case the
spatial dependence of $\check{G}_{N}(x)$ in DN is determined by the static
Usadel equation \cite{Usadel},

\begin{equation}
D \frac{\partial }{\partial x} [\check{G}_{N} (x) \frac{\partial \check{G}%
_{N}(x) }{\partial x} ] + i [\check{H},\check{G}_{N}(x)] =0,
\end{equation}
with the diffusion constant $D$ in DN, where $\check{H}$ is given by
\[
\check{H}= \left(
\begin{array}{cc}
\hat{H}_{0} & 0 \\
0 & \hat{H}_{0}%
\end{array}
\right),
\]
with $\hat{H}_{0}=\varepsilon \hat{\tau}_{3}. $

The boundary condition for $\check{G}_{N}(x)$ at the DN/S interface is given
by,
\begin{equation}
\frac{L}{R_{d}}[\check{G}_{N}(x)\frac{\partial \check{G}_{N}(x)}{\partial x}%
]_{\mid x=0_{-}}=-\frac{h}{2e^{2}R_{b}}<\check{I}>.  \label{Nazarov}
\end{equation}%
The average over the various angles of injected particles at the interface
is defined as
\begin{equation}
<\check{I}(\phi )>=\int_{-\pi /2}^{\pi /2}d\phi \cos \phi \check{I}(\phi
)/\int_{-\pi /2}^{\pi /2}d\phi T(\phi )\cos \phi
\label{average}
\end{equation}
with $\check{I}(\phi )=\check{I}$ and $T(\phi )=T_{n}$. The resistance 
of the interface $R_{b}$ is given by
\[
R_{b}=\frac{h}{2e^{2}}\frac{2}{\int_{-\pi /2}^{\pi /2}d\phi T(\phi )\cos
\phi }.
\]%
$\check{G}_{N}(-L)$ coincides with that in the normal state. The electric
current is expressed using $\check{G}_{N}(x)$ as
\begin{equation}
I_{el}=\frac{-L}{4eR_{d}}\int_{0}^{\infty }d\varepsilon \mathrm{Tr}[\tau
_{3}(\check{G}_{N}(x)\frac{\partial \check{G}_{N}(x)}{\partial x})^{K}],
\end{equation}%
where $(\check{G}_{N}(x)\frac{\partial \check{G}_{N}(x)}{\partial x})^{K}$
denotes the Keldysh component of $(\check{G}_{N}(x)\frac{\partial \check{G}%
_{N}(x)}{\partial x})$. %
In the actual calculation, we introduce a parameter $\theta (x)$ which is a
measure of the proximity effect in DN where we denoted $\theta (0)=\theta
_{0}$ in the previous subsections. Using $\theta (x)$, $\hat{R}_{N}(x)$ can
be denoted as
\begin{equation}
\hat{R}_{N}(x)=\hat{\tau}_{3}\cos \theta (x)+\hat{\tau}_{2}\sin \theta (x).
\end{equation}%
$\hat{A}_{N}(x)$ and $\hat{K}_{N}(x)$ satisfy the following equations, $\hat{%
A}_{N}(x)=-\tau _{3}\hat{R}_{N}^{\dagger }(x)\hat{\tau}_{3}$, and $\hat{K}%
_{N}(x)=\hat{R}_{N}(x)\hat{f}_{1}(x)-\hat{f}_{1}(x)\hat{A}_{N}(x)$ with the
distribution function $\hat{f}_{1}(x)$ which is given by $\hat{f}%
_{1}(x)=f_{0N}(x)+\hat{\tau}_{3}f_{3N}(x)$. In the above, $f_{3N}(x)$ is the
relevant distribution function which determines the conductance of the
junction we are now concentrating on. From the retarded or advanced
component of the Usadel equation, the spatial dependence of $\theta (x)$ is
determined by the following equation
\begin{equation}
D\frac{\partial ^{2}}{\partial x^{2}}\theta (x)+2i\varepsilon \sin [\theta
(x)]=0,  \label{Usa1}
\end{equation}%
%
%
%
%
%
while for the Keldysh component we obtain
\begin{equation}
D\frac{\partial }{\partial x}[\frac{\partial f_{3N}(x)}{\partial x}\mathrm{%
cosh^{2}}\theta _{imag}(x)]=0.  \label{Usa2}
\end{equation}%
%
%
%
%
%
with $\theta _{imag}(x)=\mathrm{Imag}[\theta (x)]$. At $x=-L$, since DN is
attached to the normal electrode, $\theta (-L)$=0 and $f_{3N}(-L)=f_{t0}$ is
satisfied with
\[
f_{t0}=\frac{1}{2}\{\tanh [(\varepsilon +eV)/(2k_{B}T)]-\tanh [(\varepsilon
-eV)/(2k_{B}T)]\},
\]%
where $V$ is the applied bias voltage.
Next, we focus on the boundary condition at the DN/S interface. Taking the
retarded part of Eq.~(\ref{Nazarov}), we obtain
\begin{equation}
\frac{L}{R_{d}}\frac{\partial \theta (x)}{\partial x}\mid _{x=0_{-}}=\frac{%
<F>}{R_{b}},  \label{prox1}
\end{equation}%
\[
F=\frac{2T_{n}[\cos \theta _{0}(f_{+}+f_{-})-\sin \theta _{0}(g_{+}+g_{-})]}{%
(2-T_{n})(1+g_{+}g_{-}+f_{+}f_{-})+T_{n}[\cos \theta _{0}(g_{+}+g_{-})+\sin
\theta _{0}(f_{+}+f_{-})]}.
\]%
On the other hand, from the Keldysh part of Eq.~(\ref{Nazarov}), we obtain
\begin{equation}
\frac{L}{R_{d}}(\frac{\partial f_{3N}}{\partial x})\mathrm{{\cosh ^{2}}}%
\theta _{0}\mid _{x=0_{-}}=-\frac{<I_{b0}>f_{3N}(0_{-})}{R_{b}},  \label{b2}
\end{equation}

\[
I_{b0}= \frac{T_{n}}{2} \frac{C_{0}f_{3N}(0_{-})}{ \mid
(2-T_{n})(1+g_{+}g_{-}+f_{+}f_{-}) +T_{n}[\cos\theta_{0} (g_{+} +g_{-}) +
\sin\theta_{0}(f_{+} + f_{-})] \mid^{2}}
\]
\[
C_{0} =T_{n}(1+\mid \cos\theta_{0} \mid^{2} + \mid \sin\theta_{0} \mid^{2})
\]
\[
\times [\mid g_{+} +g_{-} \mid^{2} + \mid f_{+} +f_{-} \mid^{2} + \mid 1+
f_{+}f_{-} + g_{+}g_{-} \mid^{2} + \mid f_{+}g_{-} - g_{+}f_{-} \mid^{2} ]
\]
\[
+ 2(2-T_{n})\mathrm{Real} \{(1+g_{+}^{*}g_{-}^{*}+f_{+}^{*}f_{-}^{*})
[(\cos\theta_{0} + \cos \theta^{*}_{0})(g_{+} + g_{-}) + (\sin\theta_{0} +
\sin \theta^{*}_{0})(f_{+} + f_{-}) ] \}
\]
\[
+ 4T_{n}\mathrm{Imag}(\cos\theta_{0} \sin\theta^{*}_{0}) \mathrm{Imag}%
[(f_{+}+f_{-})(g_{+}^{*} + g_{-}^{*})].
\]
After a simple manipulation, we can obtain $f_{3N}(0_{-})$
\[
\displaystyle f_{3N}(0_{-})=\frac{R_{b}f_{t0}}{R_{b}+\frac{R_{d}<I_{b0}>}{L}%
\int_{-L}^{0}\frac{dx}{\cosh ^{2}\theta _{imag}(x)}}
\]%
Since the electric current $I_{el}$ can be expressed via $\theta _{0}$ in
the following form%
\[
I_{el}=-\frac{L}{eR_{d}}\int_{0}^{\infty }(\frac{\partial f_{3N}}{\partial x}%
)\mid _{x=0_{-}}\cosh ^{2}[\mathrm{Imag}(\theta _{0})]d\varepsilon ,
\]%
%
%
%
%
%
we obtain the following final result for the current

\begin{equation}
I_{el}=\frac{1}{e}\int_{0}^{\infty }d\varepsilon \frac{f_{t0}}{\frac{R_{b}}{%
<I_{b0}>}+\frac{R_{d}}{L}\int_{-L}^{0}\frac{dx}{\cosh ^{2}\theta _{imag}(x)}}%
.
\end{equation}%
Then the total resistance $R$ at zero temperature is given by
\begin{equation}
R=\frac{R_{b}}{<I_{b0}>}+\frac{R_{d}}{L}\int_{-L}^{0}\frac{dx}{\cosh
^{2}\theta _{imag}(x)}.
\end{equation}
In the following section, we will discuss the normalized
conductance $\sigma _{T}(eV)=\sigma _{S}(eV)/\sigma _{N}(eV)$ where $\sigma
_{S(N)}(eV)$ is the voltage-dependent conductance in the superconducting (normal)
state given by $\sigma _{S}(eV)=1/R$ and $\sigma _{N}(eV)=\sigma
_{N}=1/(R_{d}+R_{b})$, respectively.

It should be remarked that in the present circuit theory, $R_{d}/R_{b}$ can
be varied independently of $T_{n},i.e.$ of $Z$, since we can change the
magnitude of the constriction area independently. In the other words, $%
R_{d}/R_{b}$ is no more proportional to $T_{av}(L/l)$, where $T_{av}$ is the
averaged transmissivity and $l$ is the mean free path in the diffusive
region, respectively. Based on this fact, we can choose $R_{d}/R_{b}$ and $Z$
as independent parameters. 

\section{Results}

In this subsection, we focus on the line shapes of the  conductance
where $d$-wave symmetry is chosen as a pairing symmetry of unconventional
superconductor. The pair potentials $\Delta_{\pm}$ are given by $%
\Delta_{\pm}=\Delta_{0}\cos[2(\phi \mp \alpha)]$ where $\alpha$ denotes the
angle between the normal to the interface and the crystal axis of $d$-wave
superconductors and $\Delta_{0}$ is the maximum amplitude of the pair
potential. In the above, $\phi$ denotes the injection angle of the
quasiparticle measured form the $x$-axis. It is known that quasiparticles
with injection angle $\phi$ with $\pi/4 -\mid \alpha \mid < \mid \phi \mid <
\pi/4 + \mid \alpha \mid $ feel the MABS at the interface which induces ZBCP.

\subsection{$\protect\alpha=0$ without MABS}

Let us first choose $\alpha=0$ where ZBCP due to the MABS is absent. We
choose relatively strong barrier $Z=10$ (Fig. 1) for various $R_{d}/R_{b}$.
For $E_{Th}=\Delta _{0}$ [see Fig. 1(a)], the magnitude of $\sigma _{T}(eV)$
for $\mid eV\mid <\Delta _{0}$ increases with the increase of $R_{d}/R_{b}$.
First, the line shape of the voltage-dependent
conductance remains to be $V$ shaped
and only the height of the bottom value is enhanced (curve $b$ and $c$).
The $V$ shaped line shape originates from the 
existence of nodes of the $d$-wave pair potential. 
Then, with a further increase of $R_{d}/R_{b}$, a rounded bottom structure
appears (curves $d$ and $e$). 
For $E_{Th}=0.01\Delta _{0}$ [Fig. 1(b)], the magnitude of $%
\sigma _{T}(eV)$ has a ZBCP once the magnitude of $R_{d}/R_{b}$ deviates
slightly from 0. The order of the magnitude of the ZBCP width is given by 
$E_{Th}$ as in the case of $s$-wave junctions \cite{Golubov2003}. When the
magnitude of $R_{d}/R_{b}$ exceeds unity, the $\sigma _{T}(eV)$ acquires a
zero bias conductance dip (ZBCD) (curve $e$). 
The qualitative features of line shapes of $\sigma_{T}(eV)$ is different 
from those in $s$-wave junctions (see Figs. 1 and 2 in Ref. \cite{Golubov2003}). 
%
It should be remarked that
even in the case of $d$-wave junctions we can expect ZBCP by CAR as in the
case of $s$-wave junction for $\alpha=0$.

\begin{figure}[bh]
\begin{center}
\scalebox{0.4}{
\includegraphics[width=15.0cm,clip]{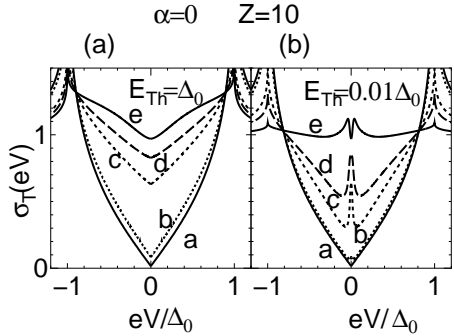}}
\end{center}
\caption{ Normalized  conductance $\sigma_{T}(eV)$ for Z=10, and $\protect\alpha=0$.
(a)$E_{Th}=\Delta_{0}$. (b)$E_{Th}=0.01\Delta_{0}$. a, $R_{d}/R_{b}=0$; b, $%
R_{d}/R_{b}=0.1$; c, $R_{d}/R_{b}=1$; d, $R_{d}/R_{b}=2$; and e, $%
R_{d}/R_{b}=10$. }
\end{figure}
On the other hand, for much more transparent case with $Z=1$ the line shapes
of the conductance become significantly different. For $E_{Th}=\Delta _{0}$
[Fig. 2(a)], the magnitude of $\sigma_{T}(eV)$ decreases with the increase
of the magnitude of $R_{d}/R_{b}$ where the bottom parts of all curves are $%
V $ shaped structures. On the other hand, for $E_{Th}=0.01\Delta _{0}$, $%
\sigma _{T}(eV)$ has a ZBCD for $R_{d}/R_{b}>1$.
Both for Figs. 2(a) and 2(b), the magnitude of $\sigma _{T}(eV)$ for $\mid
eV\mid <\Delta _{0}$ decreases with the increase of $R_{d}/R_{b}$. These
features are significantly different from those shown in Fig. 1.

\begin{figure}[bh]
\begin{center}
\scalebox{0.4}{
\includegraphics[width=15.0cm,clip]{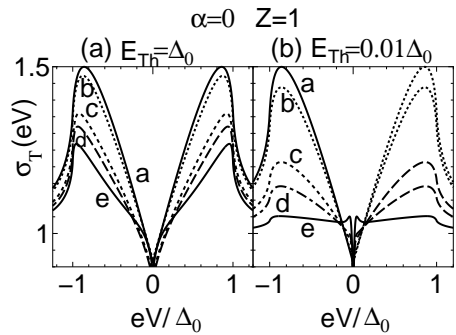}}
\end{center}
\caption{Normalized  conductance $\sigma_{T}(eV)$ for Z=1, and 
$\protect\alpha=0$. 
(a)$E_{Th}=\Delta_{0}$. (b)$E_{Th}=0.01\Delta_{0}$. a, $R_{d}/R_{b}=0$; 
b, $R_{d}/R_{b}=0.1$; c, $R_{d}/R_{b}=1$; d, $R_{d}/R_{b}=2$; and e, 
$R_{d}/R_{b}=10$.}
\label{fig:01B}
\end{figure}
%
%
For transparent limit $Z=0$, the magnitude of 
$\sigma _{T}(eV)$ decreases with the increase of $R_{d}/R_{b}$ (see Fig. 3). 
For $E_{Th}=\Delta_{0}$, $\sigma_{T}(eV)$ has a broad 
ZBCP for small $R_{d}/R_{b}$. 
However, with the increase of $R_{d}/R_{b}$, $\sigma_{T}(eV)$ for 
$\mid eV \mid <\Delta_{0}$ is reduced and becomes nearly constant. 
For $E_{Th}=0.01\Delta_{0}$ with $%
R_{d}/R_{b}=10$, tiny ZBCD appears [curve e of Fig. 3(b)] \cite{minigap}. 
As compared to
the corresponding case of $s$-wave junction (see Fig. 4 of Ref. \cite%
{Golubov2003}), ZBCD is hard to be visible in $d$-wave junctions.

\begin{figure}[tbh]
\begin{center}
\scalebox{0.4}{
\includegraphics[width=15.0cm,clip]{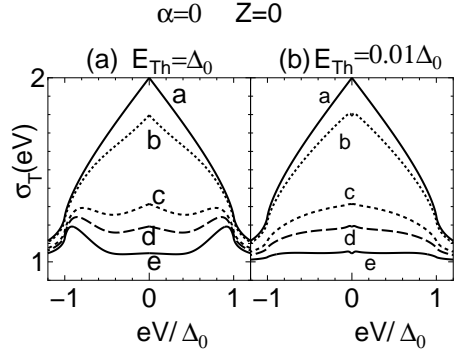}}
\end{center}
\caption{  Normalized  conductance $\sigma_{T}(eV)$ for $Z=0$, and $\protect\alpha=0$%
. (a)$E_{Th}=\Delta_{0}$. (b)$E_{Th}=0.01\Delta_{0}$. a, $R_{d}/R_{b}=0$; b,
$R_{d}/R_{b}=0.1$; c, $R_{d}/R_{b}=1$; d, $R_{d}/R_{b}=2$; and e, $%
R_{d}/R_{b}=10$. }
\label{gfig3}
\end{figure}

It is interesting to study how various parameters influence the proximity
effect. The measure of the proximity effect at the DN/US interface 
$\theta_{0}$ is plotted 
for $Z=0$ and $Z=10$ with corresponding parameters in Figs.
1 and 3 (see Figs. 4 and 5). For $R_{d}/R_{b}=0$, $\theta _{0}=0$ is
satisfied for any $E_{Th}$ and $Z$. Besides this fact, at $\varepsilon =0$, $%
\theta _{0}$ always becomes a real number. These features are consistent
with those in $s$-wave junctions \cite{Golubov2003}. First, we study the case
of $E_{Th}/\Delta _{0}=1$ (Fig. 4) where the same values of $R_{d}/R_{b}$
are chosen as in Figs. 1 and 3. The real part of $\theta _{0}$ is enhanced
with an increase in $R_{d}/R_{b}$ and decreases as a  function of $%
\varepsilon$. At the same time, the imaginary part of $\theta _{0}$ is an
increasing function of $\varepsilon $ for $\epsilon<\Delta_{0}$. 
Both real and imaginary parts have a sudden change at $\epsilon=\Delta_{0}$ 
where $\Delta_{0}$ is the magnitude of the pair potential 
felt  by quasiparticles with the perpendicular injection. 
It is remarkable that the magnitude of $\mathrm{Real(Imag)}(\theta _{0})$ is
reduced with the decrease of $Z$. 
Next, we discuss the line shapes of $%
\theta _{0}$ for $E_{Th}/\Delta _{0}=0.01$. Real($\theta _{0}$) has a peak
at zero voltage and decreases with the increase of $\varepsilon $. Imag($%
\theta _{0}$) increases sharply from 0 and has a peak at about $\varepsilon
\sim E_{Th}$, except for a sufficiently large value of $R_{d}$. 
These
features are consistent with $s$-wave junctions (see Fig. 7 of Ref. \cite%
{Golubov2003}). 
Besides this, 
both real and imaginary parts have a sudden change at $\epsilon=\Delta_{0}$ 
as in the case of  $E_{Th}/\Delta _{0}=1$. 
Also in this case,
the magnitude of $\mathrm{Real(Imag)}(\theta _{0})$
is reduced with the decrease of $Z$. 
This feature can be 
qualitatively explained as follows. We concentrate on the limiting case $%
\varepsilon=0$ for the simplicity.
The magnitude of $\theta _{0}(0)=\theta _{00}$ is determined by the
following equation [see eq. (\ref{prox1})]
\begin{equation}
\frac{\theta _{00}}{R_{d}}=<F(\phi)>
=\frac{\int_{-\pi /2}^{\pi /2}\cos \phi F(\phi
)d\phi }{R_{b}\int_{-\pi /2}^{\pi /2}\cos \phi T(\phi )d\phi },
\label{integralf}
\end{equation}%
\[
F(\phi )=\frac{2T_{n}\cos \theta _{00}\delta }{2-T_{n}+T_{n}\sin \theta
_{00}\delta },\ \ T_{n}=T(\phi )
\]%
since $f_{+}=f_{-}=\delta $ and $g_{+}=g_{-}=0$ are satisfied with $\delta
=1(-1)$ for $-\pi /4<\phi <\pi /4$ $(\pi /4<\mid \phi \mid <\pi /2)$. The
sign change nature of $\delta $  originates from $d$-wave profile of the
pair potential. This sign change reduces the magnitude of the right hand
side of eq. (\ref{integralf}), and the resulting $\theta _{00}$ is small.
For the case of large magnitude of $Z$, the degree of the reduction due to the
sign change of $\delta $, $i.e.$, $F(\phi )$ is not significant. For large
magnitude of $Z$ due to the existence of the factor $T_{n}$ ($T_{n}<<1$)
proportional to $\cos ^{2}\phi $, only the small value of $\phi $ can
contribute to the integral of numerator where $\delta =1$. This is the
reason why the obtained measure of $\theta _{00}$ for $Z=0$ is much smaller
than that for $Z=10$.

Although the magnitude of $\theta _{0}$, $i.e.$ the measure of the proximity
effect, is enhanced with increasing $R_{d}/R_{b}$, its influence on $\sigma
_{T}(eV)$ is different for low and high transparent junctions. In the low
transparent junctions, the increase in the magnitude of $\theta _{0}$ by $%
R_{d}/R_{b}$ can enhance the conductance $\sigma _{T}(eV)$ for $eV\sim 0$
and produce a ZBCP, whereas in high transparent junctions the enhancement of
$\theta _{0}$ produces the ZBCD. However, the amplitude of dip is reduced
since the magnitude of $\theta _{00}$ becomes small due to the sign change
of $F(\phi )$. %
\begin{figure}[tbh]
\begin{center}
\scalebox{0.4}{
\includegraphics[width=15.0cm,clip]{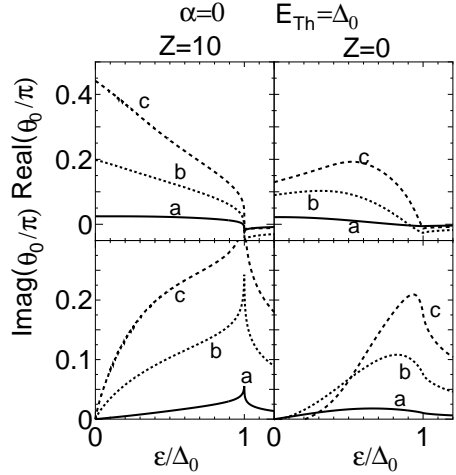}}
\end{center}
\caption{ Real(upper panels) and imaginary parts(lower panels) of $\protect%
\theta _{0}$ are plotted as a function of $\protect\varepsilon $. $Z=10$
(left panels) and $Z=0$ (right panels) with $E_{Th}/\Delta _{0}=1$ and $%
\protect\alpha =0$. a, $R_{d}/R_{b}=0.1$; b, $R_{d}/R_{b}=1$; and c, $%
R_{d}/R_{b}=10$. }
\label{gfig4}
\end{figure}

\begin{figure}[tbh]
\begin{center}
\scalebox{0.4}{
\includegraphics[width=15.0cm,clip]{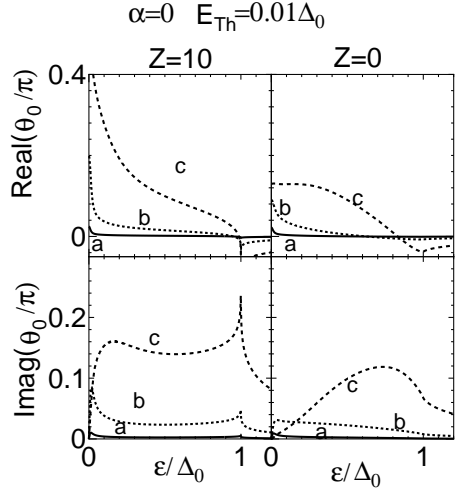}}
\end{center}
\caption{ Real(upper panels) and imaginary parts(lower panels) of $\protect%
\theta_{0}$ are plotted as a function of $\protect\varepsilon$. $Z=10$ (left
panels) and $Z=0$ (right panels) with $E_{Th}/\Delta_{0}=0.01$ and $\protect%
\alpha=0$. a, $R_{d}/R_{b}=0.1$; b, $R_{d}/R_{b}=1$; and c, $R_{d}/R_{b}=10$.
}
\label{gfig5}
\end{figure}

\subsection{ $\protect\alpha \neq 0$ with MABS}

In this subsection, we focus on $\sigma_{T}(eV)$ and $\theta_{0}$ for
$\alpha \neq 0$ $(0<\alpha<\pi/4)$. First we focus on $\alpha=\pi/8$, where MABS is formed for $%
\pi/8 < \mid \phi \mid < 3\pi/8$. In the low transparent case, $i.e.$ Z=10, $%
\sigma_{T}(eV)$ has a ZBCP due to the formation of MABS at the DN/US
interface. The height of ZBCP is reduced with the increase of $R_{d}/R_{b}$
(see Fig. 6). Contrary to the corresponding case of $\alpha=0$ (see
Fig. 1), $\sigma_{T}(eV)$ is almost independent of $E_{Th}$. For $Z=0$ (see
Fig. 7), $\sigma_{T}(eV)$ has a broad ZBCP both for $E_{Th}=\Delta_{0}$ and $%
E_{Th}=0.01\Delta_{0}$. With the increase of $R_{d}/R_{b}$, only the
magnitude of $\sigma_{T}(eV)$ is reduced and ZBCD does not appear.

\begin{figure}[tbh]
\begin{center}
\scalebox{0.4}{
\includegraphics[width=15.0cm,clip]{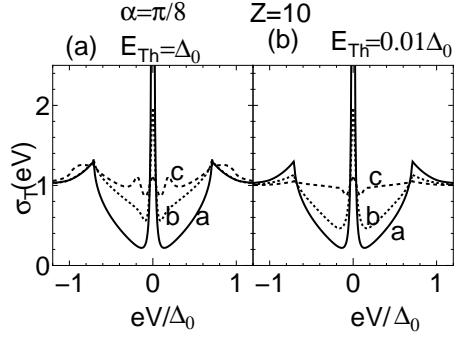}}
\end{center}
\caption{  Normalized  conductance $\sigma_{T}(eV)$ for $\protect\alpha=%
\protect\pi/8$. (a)$E_{Th}=\Delta_{0}$. (b)$E_{Th}=0.01\Delta_{0}$. a, $%
R_{d}/R_{b}=0$; b, $R_{d}/R_{b}=1$; and c, $R_{d}/R_{b}=10$. }
\label{gfig6}
\end{figure}

\begin{figure}[tbh]
\begin{center}
\scalebox{0.4}{
\includegraphics[width=15.0cm,clip]{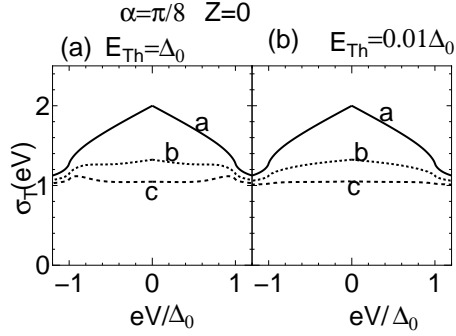}}
\end{center}
\caption{ Normalized  conductance $\sigma_{T}(eV)$
for $Z=0$, and $\protect\alpha=%
\protect\pi/8$. (a)$E_{Th}=\Delta_{0}$. (b)$E_{Th}=0.01\Delta_{0}$. a, $%
R_{d}/R_{b}=0$; b, $R_{d}/R_{b}=1$; and c, $R_{d}/R_{b}=10$. }
\label{gfig7}
\end{figure}

It is also interesting to see how $\theta_{0}$ is influenced by various 
parameters. In Fig. 8, line shapes of $\theta _{0}$ for $E_{Th}/\Delta
_{0}=1 $ are plotted for various parameters. 
$\mathrm{Real(Imag)}(\theta _{0})$ 
has a complex  $\epsilon$ dependence. 
For $Z=10$, the magnitude of $%
\mathrm{Real}(\theta _{0})$ is at $\epsilon \sim 0$ is drastically suppressed 
and is an increasing function of $\epsilon$ contrary to the case of 
$\alpha=0$. 
Both $\mathrm{Real}(\theta _{0})$ and $\mathrm{Imag}(\theta _{0})$
have a peak around $\epsilon \sim 0.7\Delta_{0} \sim \Delta_{0}\cos(2\alpha)$ 
where $\Delta_{0}\cos(2\alpha)$ is the magnitude of the pair potential 
felt by quasiparticles with perpendicular injection. 
For $Z=0$, $\varepsilon$ dependence of $%
\theta_{0}$ is qualitatively similar to the corresponding case of $\alpha=0$
(see Fig.4) since the role of MABS is not important. 
For $E_{Th}/\Delta _{0}=0.01$, the magnitude of $\mathrm{Real}%
(\theta _{0})$ at $\epsilon \sim 0$ is 
suppressed  (see Fig. 9) for $Z=10$. 
Both $\mathrm{Real}(\theta _{0})$ and $\mathrm{Imag}(\theta _{0})$
have a peak around $\epsilon \sim 0.7\Delta_{0}$
as in the case of  $E_{Th}/\Delta _{0}=1$. 
On the other hand, for high transparent case, $%
i.e.$, $Z=0$, $\mathrm{Real}(\theta _{0})$ has a 
peak at $\varepsilon=0$ and
decreases with the increase of $\varepsilon $. Imag($\theta _{0}$) increases
sharply from 0 and has a peak at about $\varepsilon \sim E_{Th}$, except for
a sufficiently large value of $R_{d}$. These features are qualitatively
consistent with $\alpha=0$. (see Fig. 5). 

%
\begin{figure}[tbh]
\begin{center}
\scalebox{0.4}{
\includegraphics[width=15.0cm,clip]{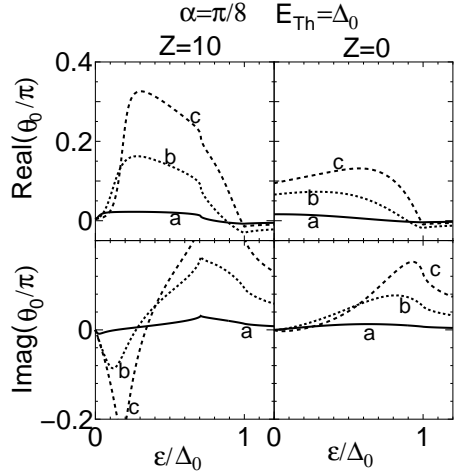}}
\end{center}
\caption{ Real(upper panels) and imaginary parts(lower panels) of $\protect%
\theta_{0}$ are plotted as a function of $\protect\varepsilon$. $Z=10$ (left
panels) and $Z=0$ (right panels) with $E_{Th}/\Delta_{0}=1$ and $\protect%
\alpha=\protect\pi/8$. a, $R_{d}/R_{b}=0.1$; b, $R_{d}/R_{b}=1$; and c, $%
R_{d}/R_{b}=10$. }
\label{gfig8c}
\end{figure}

\begin{figure}[tbh]
\begin{center}
\scalebox{0.4}{
\includegraphics[width=15.0cm,clip]{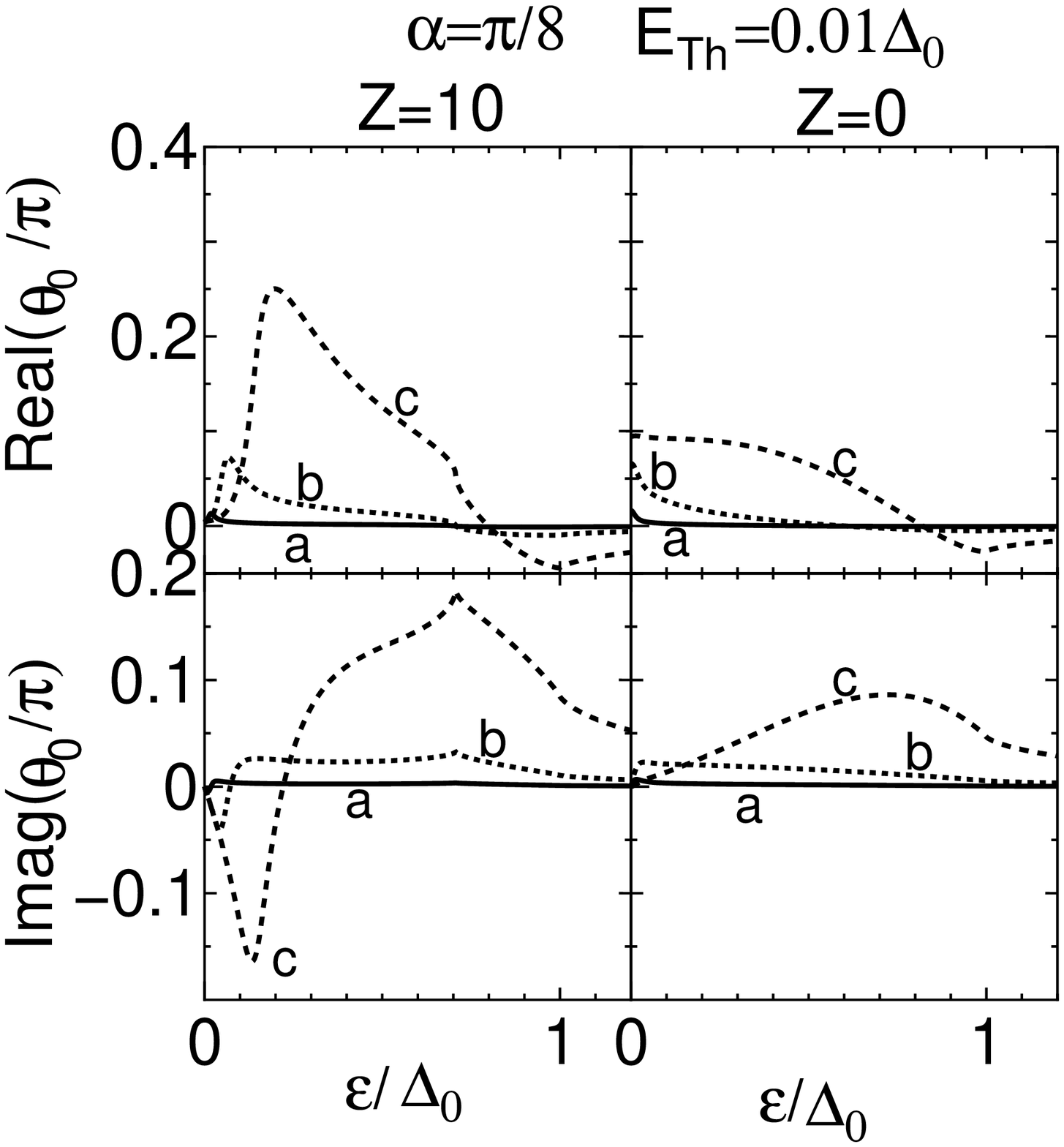}}
\end{center}
\caption{ Real(upper panels) and imaginary parts(lower panels) of $\protect%
\theta_{0}$ are plotted as a function of $\protect\varepsilon$. $Z=10$ (left
panels) and $Z=0$ (right panels) with $E_{Th}/\Delta_{0}=0.01$ and $\protect%
\alpha=\protect\pi/8$. a, $R_{d}/R_{b}=0.1$; b, $R_{d}/R_{b}=1$; and c, $%
R_{d}/R_{b}=10$. }
\label{gfig9c}
\end{figure}
In order to understand low energy profiles, we focus on the case with $%
\varepsilon=0$. $\theta_{00}$ is determined by the following relation for $%
\alpha \neq 0$.
\begin{equation}
\frac{\theta_{00}}{R_{d}} = \frac{<F(\phi)>}{R_{b}}
\end{equation}
\[
F(\phi) =
\left\{
\begin{array}{cc}
\frac{2 T(\phi) \cos \theta_{00}}{2 - T(\phi) +
T(\phi) \sin\theta_{00}} & 0< \mid \phi \mid  < \pi/4 - \alpha  \\
-2\tan\theta_{00} & \pi/4-\alpha < \mid \phi \mid < \pi/4 + \alpha \\
\frac{-2 T(\phi) \cos \theta_{00}}{2 - T(\phi) -
T(\phi) \sin\theta_{00} } &  \pi/4 + \alpha < \mid \phi \mid <\pi/2
\end{array}
\right.
\]
After simple algebra, we obtain
\begin{equation}
\frac{\theta_{00}}{R_{d}} =  \frac{f_{S1} + f_{S2} + f_{S3}} {R_{b}f_{N} }, \ \
\end{equation}
\[
f_{S1} =-2\sqrt{2}\tan \theta_{00} \sin \alpha, \ f_{S2}=\int^{\pi/4
-\alpha}_{0} \cos\phi \frac{2 T(\phi) \cos \theta_{00}}{2 - T(\phi) +
T(\phi) \sin\theta_{00}} d\phi
\]
\[
f_{S3}=-\int^{\pi/2}_{\pi/4 + \alpha} \cos\phi \frac{2 T(\phi) \cos
\theta_{00}}{2 - T(\phi) - T(\phi) \sin\theta_{00}} d\phi, \
f_{N}=\int^{\pi/2}_{0} T(\phi) \cos\phi d\phi.
\]
In the above, $f_{S1}$ denotes the contribution to $<F(\phi)>$ from the
unconventional channel with MABS and $f_{S2}$ and $f_{S3}$ denote that from
the conventional channel without MABS. For low transparent case, $T(\phi)<<1$%
, the magnitude of $f_{S1}$ dominates over those of $f_{S2}$ and $f_{S3}$
and $\theta_{00}$ is determined by
\begin{equation}
\frac{\theta_{00}}{R_{d}} \sim \frac{ -2\sqrt{2} \tan \theta_{00} \sin\alpha%
} {R_{b} f_{N}},
\end{equation}
Then the resulting $\theta_{00}$ is reduced to be almost zero. As seen from
this, MABS and proximity effect strongly competes each other. 
While for high transparent limit, $T(\phi)=1$, the magnitude
of $f_{S1}$ becomes the same order as those of $f_{S2}$ and $f_{S3}$. Then
from the contribution by conventional channel, $i.e.$ $f_{S2}$ and $f_{S3}$,
the magnitude of $\theta_{00}$ is much larger than that for $Z=10$.

For $\alpha=\pi/4$, where $f_{S2}=f_{S3}=0$ is satisfied, only the
unconventional channel $f_{S1}$ can contribute to $<F(\phi)>$. Not only $%
\theta_{00}$ but $\theta_{0}$ for any $\varepsilon$ is exactly zero. Then
the total resistance $R$ can be given by \cite{Nazarov2003}
\begin{equation}
R=\frac{R_{b}}{<I_{b0}>} + R_{d} =R_{R_{d}=0} + R_{d}
\end{equation}
and the resulting $\sigma_{T}(eV)$ is given by
\[
\sigma_{T}(eV)=\frac{ R_{d} + R_{b}}{R_{R_{d}=0} + R_{d}}
\]
One of the typical examples is plotted in Fig. 10.

\begin{figure}[tbh]
\begin{center}
\scalebox{0.4}{
\includegraphics[width=15.0cm,clip]{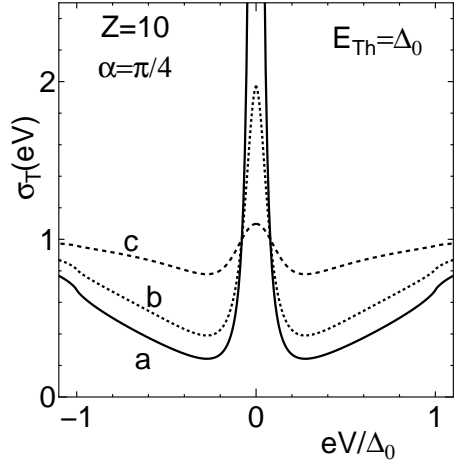}}
\end{center}
\caption{ Normalized  conductance $\sigma_{T}(eV)$
for $Z=10$, $E_{Th}/\Delta _{0}=1$
and $\protect\alpha=\protect\pi/4$. a, $R_{d}/R_{b}=0$; b, $R_{d}/R_{b}=1$; 
and c, $R_{d}/R_{b}=10$. In this case, $\protect\sigma_{T}(eV)$ is
completely independent of the magnitude of $E_{Th}$. }
\label{gfig10}
\end{figure}
The effect of $R_{d}$ is significant for the resulting $\sigma_{T}(eV)$. For
the actual quantitative comparison with tunneling experiments, we must take
into account the effect of $R_{d}$.

\subsection{$\protect\alpha$ dependence of zero-voltage conductance}

Finally, we study the dependence of $\sigma _{T}(0)$ on the angle $\alpha $%
. In this case, $\sigma _{T}(0)$ is independent of $E_{Th}$. For all
situations shown in Figs. 11 to 13, $\theta _{00}=0$ is satisfied for $%
\alpha =\pi /4$, due to the complete absence of proximity effect where only
the unconventional channel with MABS exists. For $Z=0$, $\sigma _{T}(0)$ is
almost constant with the change of $\alpha $ although $\theta _{00}$ is a
decreasing function of $\alpha $ (see Fig. 11).
\begin{figure}[tbh]
\begin{center}
\scalebox{0.4}{
\includegraphics[width=15.0cm,clip]{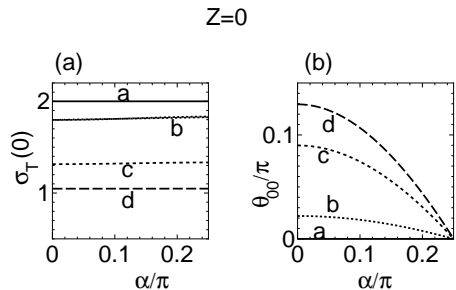}}
\end{center}
\caption{ Normalized  conductance at zero voltage $\protect\sigma %
_{T}(0)$ for various $\protect\alpha $ with $Z=0$. In this case $\protect%
\sigma _{T}(0)$ is independent of $E_{Th}$. a, $R_{d}/R_{b}=0$; b, $%
R_{d}/R_{b}=0.1$; c, $R_{d}/R_{b}=1$; and d, $R_{d}/R_{b}=10$. }
\label{gfig11}
\end{figure}

For $Z=1$, $\sigma_{T}(0)$ is an increasing function of $\alpha$, while $%
\theta_{00}$ is a decreasing function (see Fig. 12). For $Z=10$, $%
\sigma_{T}(0)$ is enhanced much more rapidly as compared to the case for $%
Z=1 $, while with the increase of $R_{d}/R_{b}$, $\sigma_{T}(0)$ becomes
nearly constant (see Fig. 13). As seen from Figs. 12 and Figs. 13, the
influence of $R_{d}/R_{b}$ on the $\sigma_{T}(0)$ is significantly important.

\begin{figure}[tbh]
\begin{center}
\scalebox{0.4}{
\includegraphics[width=15.0cm,clip]{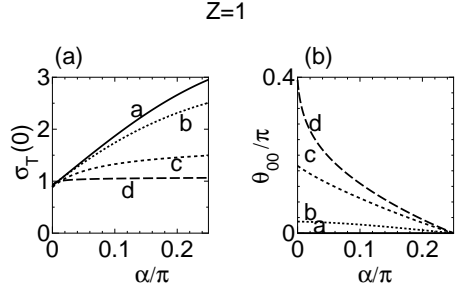}}
\end{center}
\caption{ Normalized  conductance at zero voltage $\protect\sigma%
_{T}(0)$ for various $\protect\alpha$ with $Z=1$. In this case $\protect%
\sigma_{T}(0)$ is independent of $E_{Th}$. a, $R_{d}/R_{b}=0$; b, $%
R_{d}/R_{b}=0.1$; c, $R_{d}/R_{b}=1$; and d, $R_{d}/R_{b}=10$. }
\label{gfig12}
\end{figure}

\begin{figure}[tbh]
\begin{center}
\scalebox{0.4}{
\includegraphics[width=15.0cm,clip]{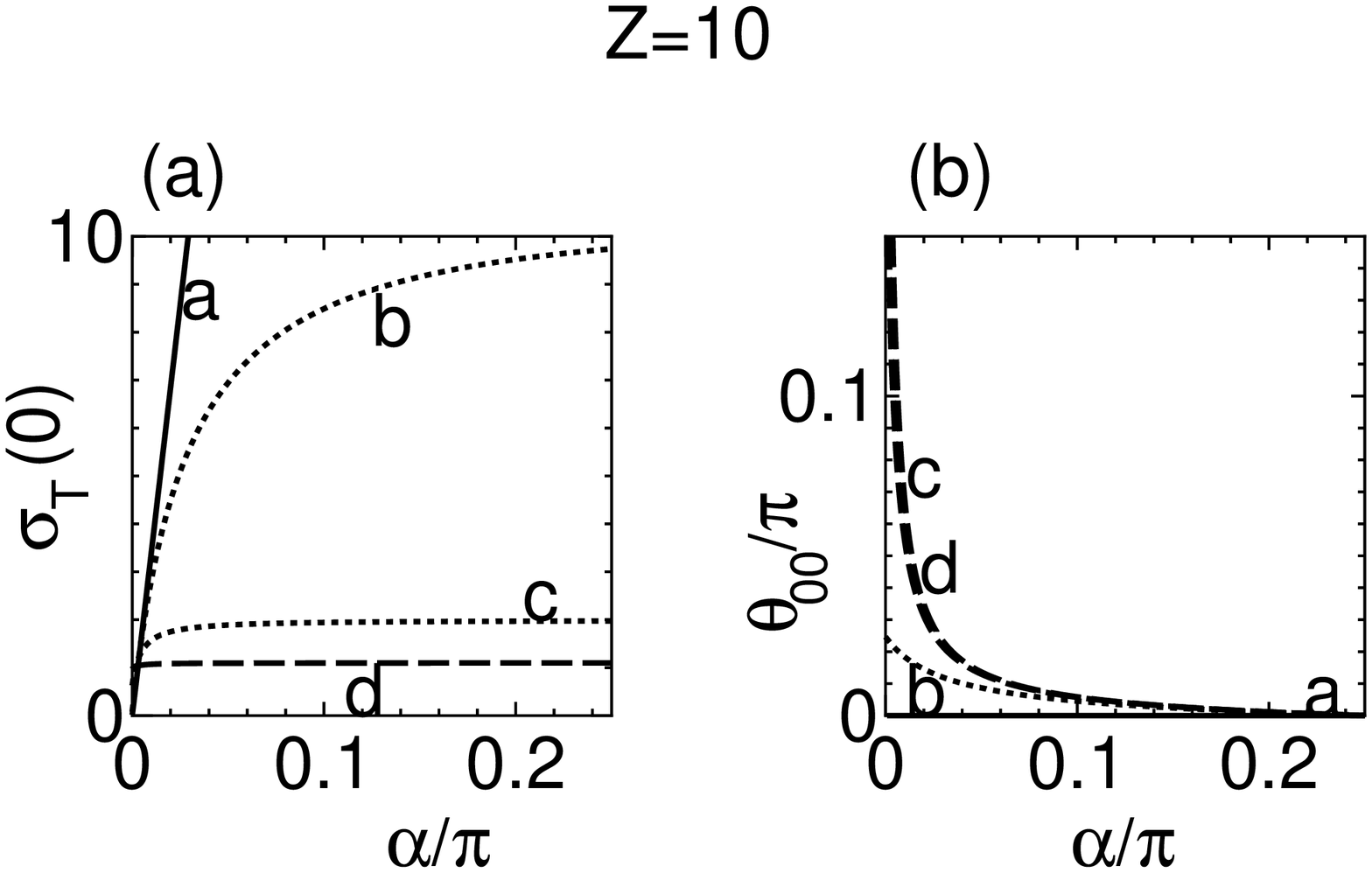}}
\end{center}
\caption{ Normalized conductance at zero voltage $\protect\sigma%
_{T}(0)$ for various $\protect\alpha$ with $Z=10$. In this case $\protect%
\sigma_{T}(0)$ is independent of $E_{Th}$. a, $R_{d}/R_{b}=0$; b, $%
R_{d}/R_{b}=0.1$; c, $R_{d}/R_{b}=1$; and d, $R_{d}/R_{b}=10$. }
\label{gfig13}
\end{figure}

In order to understand these features we look at $I_{b0}$ in detail. In
general for $\varepsilon=0$, $<I_{b0}>$ can be expressed by
\begin{equation}
<I_{b0}> =\frac{I_{b1} + I_{b2} +I_{b3}}{I_{n}}
\end{equation}
\[
I_{b1}= 2\sqrt{2} \sin \alpha \mathrm{\sec}^{2} \theta_{00}, \
I_{b2}=\int^{\pi/4 -\alpha}_{0} \cos\phi \frac{2 T(\phi)[T(\phi)
+\{2-T(\phi)\} \sin\theta_{00}] } {\mid 2 - T(\phi) + T(\phi)
\sin\theta_{00} \mid^{2}},
\]
\[
I_{b3}=\int^{\pi/2}_{\pi/4 + \alpha} \cos\phi \frac{2 T(\phi)[T(\phi) -
\{2-T(\phi)\} \sin\theta_{00}] } {\mid 2 - T(\phi) - T(\phi) \sin\theta_{00}
\mid^{2}}, \ \ I_{n}=\int^{\pi/2}_{0} T(\phi) \cos\phi d\phi.
\]
$I_{b1}$ denotes the contribution from the unconventional channel while $%
I_{b2}$ and $I_{b3}$ denote those from the conventional channel. For $Z=0$,
the integral can be performed analytically. For $\alpha=0$, $<I_{b0}>$
becomes $2\mathrm{sec}^{2}\theta_{00}[1-(\sqrt{2}-1)\sin\theta_{00}]$ for $%
\alpha=0$ and $2\mathrm{sec}^{2}\theta_{00}$ for $\alpha=\pi/4$. Since the
order of $\theta_{00}$ is at most 0.1 as seen from Fig.11(b), the difference
between the two cases is small. Then we can expect that $\sigma_{T}(0)$ is
almost constant as a function of $\alpha$. On the other hand, for $Z \neq 0$%
, the contribution from the unconventional channel becomes significant and
the resulting $<I_{b0}>$ can be approximated to be
\[
<I_{b0}> \sim 2\sqrt{2}\sin\alpha \mathrm{sec}^{2} \theta_{00}/I_{n}
\]
It is an increasing function of $\alpha$, and the resulting $\sigma_{T}(0)$
is also an increasing function of $\alpha$ (see Figs 12 and 13).


\section{Conclusions}

In the present paper, detailed theoretical investigation of the
voltage-dependent conductance of diffusive normal metal / unconventional superconductor
(DN/US) junctions is presented. We have provided the detailed derivation of
the expression for the matrix current presented in our previous paper \cite%
{Nazarov2003}. For the reader's convenience, we explicitly present
the retarded and the Keldysh parts of the matrix current for the
case when the US has a singlet parity. Applying these expressions
to DN/d-wave junctions, we have obtained the following main
results.

\noindent 1. There are two kinds of ZBCP, i.e. ZBCP due to the
coherent Andreev reflection (CAR) by proximity effect in DN and
that due to the formation of midgap Andreev bound states (MABS) at
interfaces of $d$-wave
superconductors. ZBCP frequently appears in the line shapes of $%
\sigma_{T}(eV)$. For low transparent junctions with small Thouless energy $%
E_{Th}$ we always expect ZBCP independent of $\alpha$  
except for the large magnitude of $R_{d}/R_{b}$.

\noindent 2. The nature of ZBCP due to the MABS and that by CAR is 
significantly different. 
The corresponding $\sigma _{T}(0)$ for the former case can take
arbitrary values exceeding unity. On the other hand, $\sigma _{T}(0)$ for
the latter case never exceeds unity. The width of the ZBCP in the former
case is determined by the transparency of the junction while the width for the
latter case is determined by the Thouless energy. These two ZBCPs compete
each other since the proximity effect and the existence of MABS are
incompatible in singlet junctions \cite{Nazarov2003}.

\noindent 3. For the extreme case, $\alpha=\pi/4$, where the proximity
effect is absent and the CAR is cancelled, $\sigma_{T}(eV)$ is given by $%
\sigma_{T}(eV)=(R_{b} + R_{d})/(R_{R_{d}=0} + R_{d})$ with the resistance at
the interface $R_{b}$. 
\noindent 4. Only when $\alpha=0$, MABS is 
absent for $R_{d}=0$. Then CAR influences significantly
on $\sigma_T(eV)$, similarly to the case of s-wave junction. 
When the transparency of the junction
is sufficiently low, $\sigma_{T}(eV)$ for $\mid eV \mid < \Delta_{0}$ is
enhanced with the increase of $R_{d}$ due to the enhancement of the
proximity effect assisted by CAR. 
The ZBCP becomes prominent for $E_{Th}<<\Delta_{0}$ and $%
R_{d}/R_{b}<1$. In such a case, with a further increase of $R_{d}/R_{b}$ the
ZBCP changes into a zero bias conductance dip (ZBCD).

\noindent 5. We have clarified various line shapes of the
conductance including ZBCP. The obtained results serve as an important guide
to analyze the actual experimental data of the tunneling spectra of high
$T_{C}$ cuprate junctions. We want to stress that the height of ZBCP is
strongly suppressed by the existence of DN and the resulting $\sigma _{T}(0)$
is not so high as obtained in the ballistic regime \cite{TK95}. In the
actual fit of the experimental data, we strongly hope to take into account
the effect of $R_{d}$. When the transparency of the junction is low and $%
\alpha \neq 0$, the contribution of unconventional channel becomes important
and that from conventional channel is negligible. In such a case without
solving Usadel equation $\sigma _{T}(eV)$ can be simply approximated by
\begin{equation}
\sigma _{T}(eV)=\frac{R_{d}+R_{b}}{R_{R_{d}=0}+R_{d}},\ \ R_{R_{d}=0}=\frac{%
R_{b}}{<I_{b0}>}
\end{equation}%
\[
<I_{b0}>=\frac{\int_{-\pi /2}^{\pi /2}\cos \phi I_{b0}  d\phi }{\int_{-\pi/2}^{\pi /2}\cos \phi T(\phi )d\phi },\ \ I_{b0}=\frac{ T(\phi)\{
1+T(\phi )\mid \Gamma
_{+}\mid ^{2}+[T(\phi )-1]\mid \Gamma _{+}\Gamma _{-}\mid ^{2} \}
}{\mid
1+[T(\phi )-1]\Gamma _{+}\Gamma _{-}\mid ^{2}}
\]%
\[
\Gamma _{+}=\frac{\Delta _{+} }{\varepsilon +\sqrt{\varepsilon
^{2}-\Delta _{+}^{2}}},\ \Gamma _{-}=\frac{\Delta _{-}}{\varepsilon +\sqrt{%
\varepsilon ^{2}-\Delta _{-}^{2}}},\ R_{b}=\frac{h}{2e^{2}}\frac{2}{%
\int_{-\pi /2}^{\pi /2}d\phi T(\phi )\cos \phi },
\]%
with $\Delta _{\pm }=\Delta _{0}\cos [2(\phi \mp \alpha )]$ and $\varepsilon
=eV$. This expression is a convenient one for the fit of the experimental
data. However, for the quantitative discussions including much more general
cases, one must solve Usadel equation as was done in the present paper. 
It is an interesting future problem to compare the present results 
with experiments since recent experimental results show the existence of the 
mesoscopic coherence in  high $T_{C}$ cuprate junctions \cite{Hiromi}. \par

There are several problems which are not discussed in the present paper. In
the present study, we have focused on N/S junctions. The extension of
the circuit theory to long diffusive S/N/S junctions has been
performed by Bezuglyi \textit{et al} \cite{Bezuglyi}. In S/N/S junctions,
the mechanism of multiple Andreev reflections produces the subharmonic gap
structures on I-V curves \cite{M1,M2,M3,M4,M5,M6,M7} and the situation
becomes much more complex as compared to N/S junctions. Moreover, in S/N/S
junctions with unconventional superconductors, MABS leads to the anomalous
current-phase relation and temperature dependence of the Josephson current
\cite{TKJ}. An interesting problem is an extension of the circuit theory to
S/N/S junctions with unconventional superconductors.

There are two kinds of ZBCPs considered in the present paper. We expect that
the response to the magnetic field should be significantly different in
these two cases. The ZBCP originating from MABS is rather robust against the magnetic
field while that from CAR is much more sensitive. We want to clarify this
feature in actual calculations.

In the present paper, since we follow the quasiclassical Green's function
formalism, the impurity scattering is taken into account within the
self-consistent Born approximation. It is a challenging problem to study the
weak localization effects. 

%
The authors appreciate useful and fruitful discussions with J. Inoue, H.
Itoh, Y. Asano and I. Shigeta. 
This work was supported by the Core Research for Evolutional Science
and Technology (CREST) of the Japan Science and Technology Corporation
(JST). The computational aspect of this work has been performed at the
facilities of the Supercomputer Center, Institute for Solid State Physics,
University of Tokyo and the Computer Center.
%

\section{Appendix}

The matrix current is expressed as

\begin{equation}
\check{I}=\frac{2e^{2}}{h}\mathrm{Tr}_{n,\sigma }[\bar{\Sigma}^{z}\bar{g}%
_{1}]=\frac{2e^{2}}{h}\mathrm{Tr}_{n,\sigma }[\bar{\Sigma}^{z}\bar{g}_{2}].
\end{equation}%
To find the matrix current, we have to evaluate $\bar{g}_{1(2)}$. For this
purpose, we shall consider the behavior of $\check{G}_{n}^{\sigma \sigma ^{\prime
}}(x,x^{\prime })$ in the isotropization zone in DN side $%
(-L_{1}<x,x^{\prime }<-L_{2})$ and in the ballistic zone of right side $%
(x,x^{\prime }>0)$. In the isotropic zone, since the effect of impurity
scattering is dominant and $\check{H}$ can be approximated to be $-\check{%
\Sigma}_{imp}(x)=\check{G}_{1}/(2\tau _{imp})$ for $x<0$. $\check{G}_{1}$ is
the Keldysh-Nambu Green's function in DN at $x=-L_{1}$ with $\xi
_{1}>>L_{1}>>v\tau _{imp}$ and $\xi_{1}>>L_{1}-L_{2}>>v\tau _{imp}$
where $\xi _{1}=\sqrt{D/2\pi T}$ is the coherence
length of the Green's function in DN. Due to this condition, $\check{G}_{1}$
can be approximated to be $\check{G}_{1}=\check{G}_{N}(-L_{1}) \sim \check{G}_{N}(0_{-})$, where $%
\check{G}_{N}(x)$ obeys the Usadel equation in DN. The Green's function $%
\check{G}_{n}^{\sigma \sigma ^{\prime }}(x,x^{\prime })$ is expressed by

\[
\check{G}^{\sigma\sigma^{\prime}}_{n}(x,x^{\prime}) =\bar{P}(x)(\bar{g}_{1}+%
\mathrm{sign}(x-x^{\prime})\bar{\Sigma}^{z}) \bar{P}(-x^{\prime})
\]

\begin{equation}
\bar{P}(x)=\frac{1}{2\sqrt{2v_{n}i}} \{ \exp[x/(2v_{n}\tau_{imp})] (\bar{1}-
\bar{\Sigma}^{z}\bar{G}_{1}) + \exp[-x/(2v_{n}\tau_{imp})] (\bar{1}+ \bar{%
\Sigma}^{z}\bar{G}_{1}) \}
\end{equation}
with

\begin{equation}
\bar{G}_{1}= \left(
\begin{array}{cc}
\check{G}_{1} & 0 \\
0 & \check{G}_{1}
\end{array}
\right ), \\
\bar{\Sigma}^{z} = \left (
\begin{array}{cc}
\check{1} & 0 \\
0 & -\check{1}
\end{array}
\right ). \\
\end{equation}
To ensure that $\check{G}^{\sigma\sigma^{\prime}}_{n}(x,x^{\prime})$ does
not grow with decreasing $x$ and $x^{\prime}$ in the isotropic zone in DN, we shall
require
\begin{equation}
(\bar{\Sigma}^{z} + \bar{G}_{1})(\bar{\Sigma}^{z} - \bar{g}_{1})=0,
\label{Nazarov1}
\end{equation}
\begin{equation}
(\bar{\Sigma}^{z} + \bar{g}_{1})(\bar{\Sigma}^{z} - \bar{G}_{1})=0.
\end{equation}
On the other hand, in the US side $(x>0)$ it is not simple to obtain $\check{%
G}^{\sigma\sigma^{\prime}}_{n}(x,x^{\prime})$ since it has directional
dependence. $\check{G}^{\sigma\sigma^{\prime}}_{n}(x,x^{\prime})$ is given
by
\[
\check{G}^{\sigma\sigma^{\prime}}_{n}(x,x^{\prime})= \bar{P}(x)(\bar{g}_{2}+%
\mathrm{sign}(x-x^{\prime})\bar{\Sigma}^{z}) \bar{P}(-x^{\prime}),
\]
\begin{equation}
\bar{P}(x)=\bar{P}_{1}(x)+\bar{P}_{2}(x),
\end{equation}
\[
\bar{P}_{1(2)}(x)= 
\left(
\begin{array}{cc}
\check{P}_{1(2)+} & \check{0} \\
\check{0} & \check{P}_{1(2)-}%
\end{array}
\right),
\]
\[
\check{P}_{1\pm}(x)= \frac{1}{2\sqrt{2v_{n}i}}
\left(
\begin{array}{cc}
\gamma_{\pm}(1 \mp \hat{R}_{2\pm}) & f_{0}[\gamma_{\pm}(1 \mp \hat{R}%
_{2\pm})-\gamma_{\pm}^{*}(1 \mp \hat{A}_{2\pm})] \\
0 & \gamma_{\pm}^{*}(1 \mp \hat{A}_{2\pm})%
\end{array}
\right),
\]
\[
\check{P}_{2\pm}(x)= \frac{1}{2\sqrt{2v_{n}i}}\left(
\begin{array}{cc}
\bar{\gamma}_{\pm}(1 \pm \hat{R}_{2\pm}) & f_{0}
[\bar{\gamma}_{\pm}(1 \pm \hat{R}%
_{2\pm}) -\bar{\gamma}_{\pm}^{*}(1 \pm \hat{A}_{2\pm})] \\
0 & \bar{\gamma}_{\pm}^{*}(1 \pm \hat{A}_{2\pm})%
\end{array}
\right),
\]
with
\[
\gamma_{\pm}= \exp[-i\frac{\sqrt{(\varepsilon+i\delta)^{2}-\Delta^{2}_{\pm}}
} {\hbar v_{Fx}}x],
\]
and
\[
\bar{\gamma}_{\pm}= \exp[i\frac{\sqrt{(\varepsilon+i\delta)^{2}-\Delta^{2}_{%
\pm}} } {\hbar v_{Fx}}x].
\]
In the above $\hat{R}_{2\pm}$, $\hat{A}_{2\pm}$ are retarded and advanced
component of Keldysh-Nambu Green's function $\check{G}_{2\pm}$ at the
interface of US where $\pm$ denotes the direction of motion along $x$-axis. $%
\check G_{2+}$ and $\check{G}_{2-}$ are given by $\check G_{2\pm}=\check{G}%
_{S\pm}(0_{+})$ where $\check{G}_{S\pm}(x)$ is a quasiclassical Green's
function in US. It does depend on the direction of motion $\sigma$. Here we
neglect the spatial dependence for simplicity and we assume $\check{G}%
_{2\pm}=\check{G}_{S\pm}(x)=\check{G}_{S\pm}(\infty)$. Since $\gamma_{\pm}$
and $\gamma_{\pm}^{*}$ are glowing functions with the increase of $x$, the
term, which includes this component should be eliminated by multiplying $(%
\bar{g}_{2}+\mathrm{sign}(x-x^{\prime})\bar{\Sigma}^{z})$. For convenience,
we denote
\[
\bar{g}_{2}+\mathrm{sign}(x-x^{\prime})\bar{\Sigma}^{z}= \left(
\begin{array}{cc}
\check{A} & \check{C} \\
\check{B} & \check{D}%
\end{array}
\right).
\]
In order to eliminate the divergence terms with the increase of $x$$(x>0)$,
\[
\left(
\begin{array}{cc}
\gamma_{+}(1-\hat{R}_{2+}) & f_{0}[\gamma_{+}(1-\hat{R}_{2+})-%
\gamma_{+}^{*}(1-\hat{A}_{2+})] \\
0 & \gamma_{+}^{*}(1-\hat{A}_{2+})%
\end{array}
\right) \left(
\begin{array}{cc}
\hat{a} & \hat{c} \\
\hat{b} & \hat{d}%
\end{array}
\right) =0, \ \ \check{A}= \left(
\begin{array}{cc}
\hat{a} & \hat{c} \\
\hat{b} & \hat{d}%
\end{array}
\right), \ \
\]
should be satisfied. Then following four equations must be satisfied for any
$x$
\[
\gamma_{+}(1-\hat{R}_{2+})\hat{a} + f_{0}[\gamma_{+}(1-\hat{R}_{2+})
-\gamma_{+}^{*}(1-\hat{A}_{2+})]\hat{b}=0,
\]
\[
\gamma_{+}^{*}(1-\hat{A}_{2+})\hat{b}=0,
\]
\[
\gamma_{+}^{*}(1-\hat{A}_{2+})\hat{d}=0,
\]
\[
\gamma_{+}(1-\hat{R}_{2+})\hat{c} + f_{0}[\gamma_{+}(1-\hat{R}_{2+})
-\gamma_{+}^{*}(1-\hat{A}_{2+})]\hat{d}=0.
\]
Thus we obtain
\[
\left(
\begin{array}{cc}
(1-\hat{R}_{2+}) & f_{0}[-\hat{R}_{2+} + \hat{A}_{2+})] \\
0 & (1-\hat{A}_{2+})%
\end{array}
\right) \left(
\begin{array}{cc}
\hat{a} & \hat{c} \\
\hat{b} & \hat{d}%
\end{array}
\right) =0.
\]
Applying the similar discussions for other components, we also obtain the
following equations,
\[
\left(
\begin{array}{cc}
\check{E}_{+} & 0 \\
0 & \check{E}_{-}%
\end{array}
\right) \left(
\begin{array}{cc}
\check{A} & \check{C} \\
\check{B} & \check{D}%
\end{array}
\right) =0,
\]
with
\[
\check{E}_{\pm}= \left(
\begin{array}{cc}
(1-\hat{R}_{2\pm}) & f_{0}[-\hat{R}_{2\pm} + \hat{A}_{2\pm})] \\
0 & (1-\hat{A}_{2\pm})%
\end{array}
\right).
\]
The resulting equation is identical to eq.(30) of Ref. \cite{Nazarov2}.
\begin{equation}
(\bar{\Sigma}^{z} - \bar{G}_{2}) (\bar{g}_{2} +  \bar{\Sigma}%
^{z})=0.  \label{Nazarov2}
\end{equation}
Applying similar discussion for the case with increasing $x^{\prime}$ ($%
x^{\prime}>0)$, we can also obtain
\[
(\bar{g}_{2} -  \bar{\Sigma}^{z}) (\bar{\Sigma}^{z} + \bar{G}%
_{2}) =0.
\]
To find explicit expression of $\bar{g}_{2}$, we multiply Eq. \ref{Nazarov1}
by $\bar{M}^{\dagger}$ from the left and by $\bar{M}$ from the right. By
introducing the matrix $\bar{Q}=\bar{M}^{\dagger}\bar{M}$, we derive
\begin{equation}
\bar{g}_{2} = (\bar{Q}\bar{G}_{2} + \bar{G}_{1})^{-1} \{ 2\bar{Q} + (\bar{G}%
_{1}-\bar{Q}\bar{G}_{2})\bar{\Sigma}^{z} \},
\end{equation}
with
\[
\bar{G}_{2} = \left (
\begin{array}{cc}
\check{G}_{2 +} & 0 \\
0 & \check{G}_{2 -}
\end{array}
\right ).
\]
To evaluate matrix current $\check{I}$, we rewrite $\bar{g}_{2}$ in the
basis composed of eigenvectors of $\bar{Q}$ and Keldysh-Nambu indices. For
each eigenvector $\vec{c}_{n}$ with eigenvalue $q_{n}>1$, the vector $\bar{%
\Sigma}^{z} \vec{c}_{n}$ is also an eigenvector of $\bar{Q}$ with eigenvalue
$q^{-1}_{n}$,
\begin{equation}
\bar{Q}= \left (
\begin{array}{cc}
q_{n} & 0 \\
0 & q_{n}^{-1}%
\end{array}
\right ), \ \ \bar{\Sigma}^{z}= \left(
\begin{array}{cc}
0 & \check{1} \\
\check{1} & 0%
\end{array}
\right).
\end{equation}
The resulting $\bar{g}_{2}$ is given by
\begin{equation}
\bar{g}_{2} = \left(
\begin{array}{cc}
q_{n}\check{H}_{+} + \check{G}_{1} & q_{n}\check{H}_{-} \\
q_{n}^{-1}\check{H}_{-} & q_{n}^{-1}\check{H}_{+} + \check{G}_{1}%
\end{array}
\right)^{-1} \left(
\begin{array}{cc}
q_{n}(2 - \check{H}_{-}) & \check{G}_{1}-q_{n}\check{H}_{+} \\
\check{G}_{1} - q_{n}^{-1}\check{H}_{+} & q_{n}^{-1}(2 - \check{H}_{-} )%
\end{array}
\right)
\end{equation}
with $\check{H}_{\pm}=(\check{G}_{2+} \pm \check{G}_{2-})/2$. Then the matrix
current $\check{I}$ can be expressed as
\[
\check{I} = \frac{2e^{2}}{h} \sum_{n} \check{I}_{n0}
\]

\[
\check{I}_{n0} = [\check{H}_{-}- (q_{n}\check{H}_{+}+\check{G}_{1})
\check{H}_{-}^{-1}(\check{H}_{+}/q_{n} + \check{G}_{1}) ]^{-1}(2-\check{H}_{-})
\]

\[
+ [\check{H}_{-}- (\check{H}_{+}/q_{n}+\check{G}_{1})\check{H}_{-}^{-1} (q_{n}%
\check{H}_{+} + \check{G}_{1}) ]^{-1}(2-\check{H}_{-})
\]

\[
+ [(\check{H}_{+}/q_{n}+\check{G}_{1}) - \check{H}_{-}(q_{n}\check{H}_{+} +
\check{G}_{1})^{-1}\check{H}_{-} ]^{-1}(\check{G}_{1}-\check{H}_{+}/q_{n})
\]

\begin{equation}
+ [(q_{n}\check{H}_{+} +\check{G}_{1}) - \check{H}_{-}(\check{H}_{+}/q_{n} +
\check{G}_{1})^{-1}\check{H}_{-} ]^{-1}(\check{G}_{1}-q_{n}\check{H}_{+})
\end{equation}

Applying the following identity,
\[
\check{H}_{+}^{2} + \check{H}_{-}^{2} =\check{1} \ \ \check{H}_{+}\check{H}%
_{-} + \check{H}_{-}\check{H}_{+}=0
\]
$\check{I}_{n0}$ can be written as follows,

\[
\check{I}_{n0} =( [\check{G}_{1},\check{H}_{-}^{-1}]q_{n} - \check{H}%
_{-}^{-1}\check{H}_{+} + \check{G}_{1}\check{H}_{-}^{-1}\check{H}_{+}%
\check{G}_{1}q_{n}^{2})^{-1}[(\check{G}_{1}-\check{H}_{-}^{-1}\check{G}%
_{1})q_{n} +\check{G}_{1}\check{H}_{-}^{-1}\check{H}_{+}\check{G}%
_{1}q_{n}^{2} ]
\]
\[
+ ( [\check{G}_{1},\check{H}_{-}^{-1}]/q_{n} - \check{H}_{-}^{-1}\check{H}%
_{+}/q_{n}^{2} + \check{G}_{1}\check{H}_{-}^{-1}\check{H}_{+}\check{G}%
_{1} )^{-1} [(\check{G}_{1}-\check{G}_{1}\check{H}_{-}^{-1})/q_{n} + \check{%
H}_{-}^{-1}\check{H}_{+}/q_{n}^{2} ]
\]
\[
+ ( [\check{G}_{1},\check{H}_{-}^{-1}]/q_{n} - \check{H}_{-}^{-1}\check{H}%
_{+} + \check{G}_{1}\check{H}_{-}^{-1}\check{H}_{+}\check{G}%
_{1}/q_{n}^{2})^{-1}[(\check{G}_{1}-\check{H}_{-}^{-1}\check{G}_{1})/q_{n}
+ \check{G}_{1}\check{H}_{-}^{-1}\check{H}_{+}\check{G}_{1}/q_{n}^{2} ]
\]
\begin{equation}
+ ( [\check{G}_{1},\check{H}_{-}^{-1}]q_{n} - \check{H}_{-}^{-1}\check{H}%
_{+}q_{n}^{2} + \check{G}_{1}\check{H}_{-}^{-1}\check{H}_{+}\check{G}%
_{1} )^{-1} [(\check{G}_{1}-\check{G}_{1}\check{H}_{-}^{-1})q_{n} + \check{H%
}_{-}^{-1}\check{H}_{+}q_{n}^{2} ]
\end{equation}
The eigenvalue $q_{n}$ and the transparency of the junction $T_{n}$ satisfy
the following relations,
\[
4q_{n}/(1 + q_{n})^{2}=T_{n}, \ \ (1 + q_{n}^{2})/(1 +
q_{n})^{2}=(2-T_{n})/2
\]
We introduce $T_{1n}$
\[
T_{1n}=\frac{T_{n}}{2-T_{n} + 2\sqrt{1-T_{n}}}
\]
and the resulting $\check{I}_{n0}$ becomes
\[
\check{I}_{n0} =-\check{D}^{-1} [T_{1n}(2\check{G}_{1}-[\check{H}_{-}^{-1},%
\check{G}_{1}]_{+} ) + \check{H}_{-}^{-1}\check{H}_{+} + T_{1n}^{2}\check{G}%
_{1}\check{H}_{-}^{-1}\check{H}_{+}\check{G}_{1} ]
\]
\begin{equation}
+ \check{G}_{1}\check{D}^{-1}\check{G}_{1} [T_{1n}(2\check{G}_{1}-[\check{H}%
_{-}^{-1},\check{G}_{1}]_{+} ) + T_{1n}^{2}\check{H}_{-}^{-1}\check{H}_{+} +
\check{G}_{1}\check{H}_{-}^{-1}\check{H}_{+}\check{G}_{1} ]
\end{equation}

with
\[
\check{D}=-T_{1n}[\check{G}_{1},\check{H}_{-}^{-1}]+\check{H}_{-}^{-1}\check{%
H}_{+}-T_{1n}^{2}\check{G}_{1}\check{H}_{-}^{-1}\check{H}_{+}\check{G}_{1}
\]%
$\check{I}_{n0}$ is also represented as follows \cite{Nazarov2003},
\[
\check{I}_{n0}=2[\check{G}_{1},\check{B}_{n}]
\]%
with
\begin{equation}
\check{B}_{n}=(-T_{1n}[\check{G}_{1},\check{H}_{-}^{-1}]+\check{H}_{-}^{-1}%
\check{H}_{+}-T_{1n}^{2}\check{G}_{1}\check{H}_{-}^{-1}\check{H}_{+}%
\check{G}_{1})^{-1}(T_{1n}(1-\check{H}_{-}^{-1})+T_{1n}^{2}\check{G}_{1}%
\check{H}_{-}^{-1}\check{H}_{+}).
\end{equation}



\begin{thebibliography}{999}
\bibitem{Andreev} A.F. Andreev, Zh. Eksp. Teor. Fiz. 
{\bf 46} 1823 [Sov. Phys. JETP \textbf{19} 1228 (1964)].

\bibitem{Hekking} F. W. J. Hekking and Yu. V. Nazarov, Phys. Rev. Lett.
\textbf{71} 1625 (1993). 

\bibitem{Giazotto} F. Giazotto, P. Pingue, F. Beltram, M. Lazzarino, D.
Orani, S. Rubini, and A. Franciosi, Phys. Rev. Lett. \textbf{87} 216808
(2001).

\bibitem{Klapwijk} T.M. Klapwijk, Physica B \textbf{197} 481 (1994).

\bibitem{Kastalsky} A. Kastalsky, A.W. Kleinsasser, L.H. Greene, R. Bhat,
F.P. Milliken, J.P. Harbison, Phys. Rev. Lett. \textbf{67} 3026 (1991).

\bibitem{Nguyen} C. Nguyen, H. Kroemer and E.L. Hu, Phys. Rev. Lett. \textbf{%
69} 2847 (1992).

\bibitem{Wees} B.J. van Wees, P. de Vries, P. Magnee, and T.M. Klapwijk,
Phys. Rev. Lett. \textbf{69} 510 (1992).

\bibitem{Nitta} J. Nitta, T. Akazaki and H. Takayanagi, Phys. Rev. B \textbf{%
49} 3659 (1994).

\bibitem{Bakker} S.J.M. Bakker, E. van der Drift, T.M. Klapwijk, H.M.
Jaeger, and S. Radelaar, Phys. Rev. B \textbf{49} 13275 (1994).

\bibitem{Xiong} P. Xiong, G. Xiao and R.B. Laibowitz, Phys. Rev. Lett.
\textbf{71} 1907 (1993).

\bibitem{Magnee} P.H.C. Magnee, N. van der Post, P.H.M. Kooistra, B.J. van
Wees, and T.M. Klapwijk, Phys. Rev. B \textbf{50} 4594 (1994).

\bibitem{Kutch} J. Kutchinsky, R. Taboryski, T. Clausen, C. B. Sorensen, A.
Kristensen, P. E. Lindelof, J. Bindslev Hansen, C. Schelde Jacobsen, and J.
L. Skov, Phys.Rev. Lett. \textbf{78} 931 (1997).

\bibitem{Poirier} W. Poirier, D. Mailly, and M. Sanquer, Phys. Rev. Lett.
\textbf{79} 2105 (1997).

%

\bibitem{BTK} G.E. Blonder, M. Tinkham, and T.M. Klapwijk, Phys. Rev. B
\textbf{25} 4515 (1985).

\bibitem{Zaitsev} A. V. Zaitsev, Zh. Eksp. Teor. Fiz. 
{\bf 86} 1724 (1984) [Sov. Phys. JETP  \textbf{59} 1163 (1984)].

%

\bibitem{Beenakker1} C.W.J. Beenakker, Rev. Mod. Phys. \textbf{69} 731
(1997);

\bibitem{Lambert} C.J. Lambert, J. Phys. Condens. Matter \textbf{3} 6579
(1991);

\bibitem{Takane} Y. Takane and H. Ebisawa, J. Phys. Soc. Jpn. \textbf{61}
2858 (1992).


\bibitem{Beenakker2} C.W.J. Beenakker, Phys. Rev. B \textbf{46} 12841 (1992).

\bibitem{reflec} C. W. J. Beenakker, B. Rejaei, and J. A. Melsen, Phys. Rev.
Lett. \textbf{72} 2470 (1994).

\bibitem{Lesovik} G.B. Lesovik, A.L. Fauchere, and G. Blatter, Phys. Rev. B
\textbf{55} 3146 (1997).


\bibitem{Larkin} A.I. Larkin and Yu. V. Ovchinnikov, 
Zh. Eksp. Teor. Fiz. \textbf{68} 1915 (1975) [Sov. Phys. JETP  
\textbf{41} 960 (1975).]

\bibitem{Volkov} A.F. Volkov, A.V. Zaitsev and T.M. Klapwijk, Physica C 210
21 (1993).

\bibitem{KL} M.Yu. Kuprianov and V. F. Lukichev, Zh. Exp. Teor. Fiz.
\textbf{94}  139 (1988)[Sov. Phys. JETP \textbf{67}  1163 (1988)]

\bibitem{Usadel} K.D. Usadel Phys. Rev. Lett. \textbf{25} 507 (1970).

\bibitem{Nazarov1} Yu. V. Nazarov, Phys. Rev. Lett. \textbf{73} 1420 (1994).

\bibitem{Yip} S. Yip Phys. Rev. B \textbf{52} 3087 (1995).

\bibitem{Stoof} Yu. V. Nazarov and T. H. Stoof, Phys. Rev. Lett. \textbf{76}%
, 823 (1996); T. H. Stoof and Yu. V. Nazarov, Phys. Rev. B \textbf{53},
14496 (1996).

\bibitem{Reentrance} A. F. Volkov, N. Allsopp, and C. J. Lambert, J. Phys.
Cond. Mat. \textbf{8}, L45 (1996); A. F. Volkov and H. Takayanagi, Phys.
Rev. B \textbf{56}, 11184 (1997).

\bibitem{Golubov} A.A. Golubov, F.K. Wilhelm, and A.D. Zaikin, Phys. Rev. B
\textbf{55}, 1123 (1997).

\bibitem{Takayanagi} A.F. Volkov and H. Takayanagi, Phys. Rev. B \textbf{56}%
, 11184 (1997).

\bibitem{Seviour} R. Seviour and A. F. Volkov, Phys. Rev. B \textbf{61},
R9273 (2000).

\bibitem{Belzig} W. Belzig, F. K. Wilhelm, C. Bruder, G. Sch\"{o}n, 
and A. D. Zaikin, 
Superlattices and Microstructures \textbf{25}, 1251 (1999).

\bibitem{minigap} A.A. Golubov and M.Yu. Kuprianov, J. Low Temp. Phys.
\textbf{70}, 83 (1988); W. Belzig, C. Bruder, and G. Sch\"{o}n, Phys. Rev. B
\textbf{54}, 9443 (1996); J.A. Melsen, P.W.Brouwer, K.M. Frahm, and C.W.J.
Beenakker, Europhys. Lett. \textbf{35}, 7, (1996); F. Zhou, P. Charlat, B.
Spivak, and B. Pannetier, J. Low Temp. Phys. \textbf{110}, 841 (1998).


\bibitem{Lambert1} C. J. Lambert, R. Raimondi, V. Sweeney and A. F. Volkov,
Phys. Rev. B \textbf{55} 6015 (1997).

\bibitem{Bezuglyi} E. V. Bezuglyi, E. N. Bratus', V. S. Shumeiko, G. Wendin
and H. Takayanagi, Phys. Rev. B \textbf{62} 14439 (2000).

\bibitem{Nazarov2} Yu. V. Nazarov, Superlatt. Microstruct. \textbf{25} 1221
(1999), cond-mat/9811155.

\bibitem{Golubov2003} Y. Tanaka, A. A. Golubov and S. Kashiwaya, Phys. Rev. B
\textbf{68} 054513 (2003).



\bibitem{Buch} L.J. Buchholtz and G. Zwicknagl, Phys. Rev. B \textbf{23}
5788 (1981); C. Bruder, Phys. Rev. B \textbf{41} 4017 (1990); C.R. Hu, Phys.
Rev. Lett. \textbf{72}, 1526 (1994).

\bibitem{TK95} Y. Tanaka and S. Kashiwaya, Phys. Rev. Lett. \textbf{74},
3451 (1995); S. Kashiwaya, Y. Tanaka, M. Koyanagi and K. Kajimura, Phys.
Rev. B \textbf{53}, 2667 (1996).

\bibitem{96} Y. Tanaka and S. Kashiwaya, Phys. Rev. B \textbf{53}, 9371 
(1996).

\bibitem{Kashi00} S. Kashiwaya and Y. Tanaka, Rep. Prog. Phys. \textbf{63},
1641 (2000) and references therein. 
%
%
%

\bibitem{e1} J. Geerk, X.X. Xi, and G. Linker: Z. Phys. B. \textbf{73}, 
329 (1988).

\bibitem{e2} S. Kashiwaya, Y. Tanaka, M. Koyanagi, H. Takashima, and K.
Kajimura, Phys. Rev. B \textbf{51} 1350 (1995).

\bibitem{e3} L. Alff, H. Takashima, S. Kashiwaya, N. Terada, H. Ihara, Y.
Tanaka, M. Koyanagi, and K. Kajimura, Phys. Rev. B {55}, 14757 (1997).

\bibitem{e4} M. Covington, M. Aprili, E. Paraoanu, L.H. Greene, F. Xu, J.
Zhu, and C.A. Mirkin, Phys. Rev. Lett. \textbf{79}, 277 (1997).

\bibitem{e5} J. Y. T. Wei, N.-C. Yeh, D. F. Garrigus and M. Strasik: Phys.
Rev. Lett. \textbf{81},  2542 (1998).

\bibitem{e6} I. Iguchi, W. Wang, M. Yamazaki, Y. Tanaka, and S. Kashiwaya:
Phys. Rev. B \textbf{62},  R6131 (2000).

\bibitem{e7} F. Laube, G. Goll, H.v. L\"{o}hneysen, M. Fogelstr\"{o}m, and
F. Lichtenberg, Phys. Rev. Lett. \textbf{84}, 1595 (2000).

\bibitem{e8} Z.Q. Mao, K.D. Nelson, R. Jin, Y. Liu, and Y. Maeno, Phys. Rev.
Lett. \textbf{87}, 037003 (2001).

\bibitem{e9} Ch. W\"{a}lti, H.R. Ott, Z. Fisk, and J.L. Smith, Phys. Rev.
Lett. \textbf{84}, 5616 (2000).

\bibitem{e10} H. Aubin, L. H. Greene, Sha Jian and D. G. Hinks, Phys. Rev.
Lett. \textbf{89}, 177001 (2002).

\bibitem{e11} Z. Q. Mao, M. M. Rosario, K. D. Nelson, K. Wu, I. G. Deac, P.
Schiffer, Y. Liu, T. He, K. A. Regan, and R. J. Cava Phys. Rev. B \textbf{67}%
, 094502 (2003).

\bibitem{e12} A. Sharoni, O. Millo, A. Kohen, Y. Dagan, R. Beck, G.
Deutscher, and G. Koren Phys. Rev. B \textbf{65}, 134526 (2002).

\bibitem{e13} A. Kohen, G. Leibovitch, and G. Deutscher Phys. Rev. Lett.
\textbf{90}, 207005 (2003).

\bibitem{e14} M. M. Qazilbash, A. Biswas, Y. Dagan, R. A. Ott, and R. L.
Greene, Phys. Rev. B \textbf{ 68}, 024502 (2003).

\bibitem{e15} J. W. Ekin, Y. Xu, S. Mao, T. Venkatesan, D. W. Face, M. Eddy,
and S. A. Wolf: Phys. Rev. B \textbf{56},  13746 (1997).

\bibitem{T1} Y.~Tanuma, Y.~Tanaka, M.~Yamashiro and S.~Kashiwaya : Phys.
Rev. B \textbf{57}, 7997 (1998).

\bibitem{T2} Y. Tanuma, Y. Tanaka, M. Ogata and S. Kashiwaya: J. Phys. Soc.
Jpn., \textbf{67},  1118 (1998).

\bibitem{T3} Y.~Tanuma, Y.~Tanaka, M. Ogata and S.~Kashiwaya: Phys. Rev. B
\textbf{60},  9817 (1999).

\bibitem{T4} Y. Tanuma, Y. Tanaka, and S. Kashiwaya: Rhys. Rev. B \textbf{64}, 
 214519 (2001).

\bibitem{T5} M.~Fogelstr\"{o}m, D.~Rainer, and J.~A.~Sauls: Phys. Rev. Lett.
\textbf{79},  281 (1997); D.~Rainer, H.~Burkhardt, M.~Fogelstr\"{o}m, and
J.~A.~Sauls: J. Phys. Chem. Solids \textbf{59}, 2040 (1998).

\bibitem{T6} M. Matsumoto and H. Shiba, J. Phys. Soc. Jpn. \textbf{64}, 
 1703 (1995); M. Matsumoto and H. Shiba, J. Phys. Soc. Jpn. \textbf{64}, 
 4867 (1995).

\bibitem{T7} L. J. Buchholtz, M. Palumbo, D. Rainer and J. A. Sauls, J. Low
Temp. Phys. \textbf{101}, 1097 (1995).

\bibitem{T8} Y.~Tanaka and S.~Kashiwaya, Phys. Rev. B \textbf{58}, 2948
(1998).

\bibitem{tr1} M.~Yamashiro, Y.~Tanaka and S.~Kashiwaya: Phys. Rev. B 
\textbf{56}, 7847 (1997).

\bibitem{tr2} M.~Yamashiro, Y.~Tanaka Y. Tanuma and S.~Kashiwaya, J. Phys.
Soc. Jpn. \textbf{67}, 3224 (1998).

\bibitem{tr3} M.~Yamashiro, Y.~Tanaka and S.~Kashiwaya, J. Phys. Soc. Jpn.
\textbf{67}, 3364 (1998).

\bibitem{tr4} M. Yamashiro, Y. Tanaka Y. Tanuma and S. Kashiwaya: J. Phys.
Soc. Jpn. \textbf{68} 2019  (1999).

\bibitem{tr5} C.~Honerkamp and M.~Sigrist, Prog. Theor. Phys. \textbf{100},
53 (1998).

\bibitem{tr6} Y. Asano, Y. Tanaka, Y. Matsuda and S. Kashiwaya: 
Phys. Rev. B 68, 184506 (2003) \par


\bibitem{or1} Y.~Tanuma,  K.~Kuroki, Y.~Tanaka, and S.~Kashiwaya: Phys. Rev. B
\textbf{64} (2001) 214510.

\bibitem{or2} K. Sengupta, I. \v{Z}uti\'c, H.-J. Kwon, V.M. Yakovenko, and
S. Das Sarma: Phys. Rev. B \textbf{63} (2001) 144531.

\bibitem{or3} Y.~Tanuma, K.~Kuroki, Y.~Tanaka, R.~Arita, S.~Kashiwaya and
H.~Aoki: Phys. Rev. B \textbf{66},  094507 (2002).

\bibitem{dop1} Y.~Tanuma, Y.~Tanaka, K.~Kuroki,  and S.~Kashiwaya: Phys. Rev.
B \textbf{66}, 174502 (2002). 

\bibitem{dop2} Y. Tanaka, H. Tsuchiura, Y. Tanuma and S. Kashiwaya: J. Phys.
Soc. Jpn. \textbf{71},  271 (2002).

\bibitem{dop3} Y. Tanaka, Y. Tanuma K. Kuroki and S. Kashiwaya: J. Phys.
Soc. Jpn. \textbf{71},  2102 (2002).

\bibitem{dop4} Y. Tanaka, H. Itoh, H. Tsuchiura, Y. Tanuma, J. Inoue, and S.
Kashiwya: J. Phys. Soc. Jpn. \textbf{71},  2005 (2002).

\bibitem{dop5} Yu. S. Barash, M. S. Kalenkov, and J. Kurkijarvi Phys. Rev. B
\textbf{62}, 6665 (2000).

\bibitem{f1} J-X. Zhu, B. Friedman, and C. S. Ting: Phys. Rev. B \textbf{59}, 
 9558 (1999).

\bibitem{f2} S. Kashiwaya, Y. Tanaka, N. Yoshida and M.R. Beasley : Phys.
Rev. B \textbf{60},  3572 (1999).

\bibitem{f3} I. Zutic and O. T. Valls: Phys. Rev. B \textbf{60}  6320 (1999).

\bibitem{f4} N. Yoshida, Y. Tanaka, J. Inoue, and S. Kashiwaya: J. Phys.
Soc. Jpn. \textbf{68}  1071 (1999).

\bibitem{f5} T. Hirai, N. Yoshida, Y. Tanaka, J. Inoue and S. Kashiwaya: J.
Phys. Soc. Jpn. \textbf{70}  1885 (2001).

\bibitem{f6} N. Yoshida, H. Itoh, T. Hirai, Y. Tanaka, J. Inoue and S.
Kashiwaya: Phsica C \textbf{367}  135 (2002).

\bibitem{f7} T. Hirai, Y. Tanaka, N. Yoshida, Y. Asano, J. Inoue and S.
Kashiwaya Phys. Rev. B \textbf{67} 174501 (2003).

\bibitem{f8} Y.~Tanaka and S.~Kashiwaya, J. Phys. Soc. Jpn. \textbf{68},
3485 (1999).

\bibitem{f9} Y.~Tanaka and S.~Kashiwaya, J. Phys. Soc. Jpn. \textbf{69},
1152 (2000).

\bibitem{TKJ} Y. Tanaka and S. Kashiwaya, Phys. Rev. B \textbf{53}, 11957
(1996); Y. Tanaka and S. Kashiwaya, Phys. Rev. B \textbf{56}, 892 (1997).

\bibitem{J1} E.~Il'ichev, V.~Zakosarenko, R.~P.~J.~IJsselsteijn,
V.~Schultze, H.~-G.~Meyer, H.~E.~Hoenig, H.~Hilgenkamp, and J.~Mannhart,
Phys. Rev. Lett. \textbf{81}, 894 (1998).

\bibitem{J2} Y.~Asano, Phys. Rev. B \textbf{63}, 052512 (2001).

\bibitem{J3} Y.~Asano, Phys. Rev. B \textbf{64}, 014511 (2001).

\bibitem{J4} Y.~Asano, Phys. Rev. B \textbf{64}, 224515 (2001).

\bibitem{J5} Y.~Asano, J. Phys. Soc. Jpn. \textbf{71}, 905 (2002).

\bibitem{J6} Y. Asano, Y. Tanaka, M. Sigrist and S. Kashiwaya, Phys. Rev. B
\textbf{67} 184505 (2003).

\bibitem{J7} Y. S. Barash, H. Burkhardt and D. Rainer, Phys. Rev. Lett.
\textbf{77}, 4070 (1996).

\bibitem{J8} E. Ilichev, M. Grajcar, R. Hlubina, 
R. P. J. IJsselsteijn, H. E. Hoenig, 
H.-G. Meyer, A. Golubov, M. H. S. Amin, A. M. Zagoskin, A. N.
Omelyanchouk and M. Yu. Kuprianov, Phys. Rev. Lett. \textbf{86}, 5369 (2001).

\bibitem{J9} G. Testa, A. Monaco, E. Esposito, E. Sarnelli, D.-J. Kang,
E.J. Tarte, S.H. Mennema and M.G. Blamire, cond-mat/0310727.

\bibitem{o1} Y. Tanaka, T. Hirai, K. Kusakabe and S. Kashiwaya: Phys. Rev. B
\textbf{60},  6308 (1999).

\bibitem{o2} T. Hirai, K. Kusakabe and Y. Tanaka Physica C \textbf{336}
 107 (2000); K. Kusakabe and Y. Tanaka Physica C \textbf{367} 123 (2002);
K. Kusakabe and Y. Tanaka; J. Phys. Chem. Solids \textbf{63} 1511 (2002).

\bibitem{o3} N. Stefanakis Phys. Rev. B \textbf{64}, 224502 (2001).

\bibitem{o4} Z. C. Dong, D. Y. Xing, and Jinming Dong Phys. Rev. B \textbf{65%
}, 214512 (2002); Z. C. Dong, D. Y. Xing, Z. D. Wang, Ziming Zheng, and
Jinming Dong Phys. Rev. B \textbf{63}, 144520 (2001).

\bibitem{o5} M. H. S. Amin, A. N. Omelyanchouk, and A. M. Zagoskin Phys.
Rev. B 63, 212502 (2001).

\bibitem{o6} Shin-Tza Wu and Chung-Yu Mou Phys. Rev. B \textbf{66}, 012512
(2002).

\bibitem{o7} A.~A.~Golubov, M.~Y.~Kuprianov: Pis'ma Zh. Eksp. Teor. fiz
\textbf{69}  242 (1999)[ Sov. Phys. JETP Lett. \textbf{69}  262 (1999)];
\textbf{67}  478 (1998)[ Sov. Phys. JETP Lett. \textbf{67}  501 (1998)]. 

\bibitem{o8} A.~Poenicke, Yu.~S.~Barash, C.~Bruder, and V.~Istyukov: Phys.
Rev. B \textbf{59}, 7102 (1999); K.~Yamada, Y.~Nagato, S.~Higashitani and
K.~Nagai: J. Phys. Soc. Jpn. \textbf{65},  1540 (1996).

\bibitem{o9} T.~L\"{u}ck, U.~Eckern, and A.~Shelankov: Phys. Rev. B \textbf{%
63}, 064510  (2002) .

\bibitem{Nazarov2003} Y. Tanaka, Y.V.  Nazarov and S. Kashiwaya, Phys. Rev.
Lett. {\bf 90},  167003 (2003).



\bibitem{triplet2003} Y. Tanaka and S. Kashiwaya, condmat[0308123].

\bibitem{Hiromi} H. Kashiwaya, A.Sawa, S. Kashiwaya, H. Yamasaki, M.
Koyanagi, I. Kurosawa, Y. Tanaka and I. Iguchi Physica C, \textbf{357-360}
1610 (2001); H. Kashiwaya, I. Kurosawa, S. Kashiwaya, A. Sawa and Y. Tanaka
Phys. Rev. B \textbf{68} 054527 (2003).





\bibitem{M1} T. M. Klapwijk, G. E. Blonder, and M. Tinkham, Physica B and C
\textbf{109-110} 1657 (1982).

\bibitem{M2} M. Octavio, M. Tinkham, G. E. Blonder, and T. M. Klapwijk,
Phys. Rev. B \textbf{27} 6739 (1983).

\bibitem{M3} G.B. Arnlod, J. Low Temp. Phys. \textbf{68} 1 (1987); U.
Gunsenheimer and A. D. Zaikin, Phys. Rev. B \textbf{50} 6317 (1994).

\bibitem{M4} E.N. Bratus', V.S. Shumeiko, and G. Wendin, Phys. Rev. Lett.
\textbf{74} 2110 (1995); D. Averin and A. Bardas, Phys. Rev. Lett. \textbf{75%
} 1831 (1995); J.C. Cuevas, A. Martin-Rodero and A. L.  Yeyati, Phys. Rev. B
\textbf{54} 7366 (1996).

\bibitem{M5} A. Bardas and D. V. Averin, Phys. Rev. B \textbf{56} 8518
(1997); A. V. Zaitsev and D. V. Averin, Phys. Rev. Lett. \textbf{80} 3602
(1998).

\bibitem{M6} A. V. Zaitsev, Physica C \textbf{185-189} 2539 (1991).

\bibitem{M7} E.V. Bezuglyi, E. N. Bratus', V. S. Shumeiko and G. Wendin,
Phys. Rev. Lett. \textbf{83} 2050 (1990).
\end{thebibliography}
\end{document}